\documentclass[fleqn,usenatbib]{mnras}
\usepackage{newtxtext,newtxmath}
\usepackage{epsfig}

\usepackage{amssymb}	
\usepackage{amsmath}	
\usepackage{natbib}
\usepackage{threeparttable} 
\usepackage{graphicx}
\usepackage{setspace}
\usepackage{longtable}
\usepackage{multicol}
\usepackage{subfigure}

\usepackage{float} 
\usepackage{graphicx}	

\usepackage{lscape}  
\usepackage{mathtools,graphicx,array} 
\usepackage{booktabs}  
\usepackage{multirow}
\usepackage{verbatim}
\usepackage{epstopdf}

\title[\rm multi-peaked bursts in 4U 1636-53]{The study of multi-peaked type-I X-ray bursts in the neutron-star low mass X-ray binary 4U 1636$-$536 with RXTE}

\author[\rm Chen Li et al.]{
Chen Li$^{1,2,3,4}$\thanks{E-mail: lichen@ynao.ac.cn},
Guobao Zhang$^{1,2,3,4}$\thanks{E-mail: zhangguobao@ynao.ac.cn},
Mariano M\'endez $^{5}$,
Jiancheng Wang$^{1,2,3,4}$,
\newauthor
and Ming Lyu$^{6,7}$
\\
$^{1}$Yunnan Observatories, Chinese Academy of Sciences, Kunming 650216, People's Republic of China \\
$^{2}$Key Laboratory for the Structure and Evolution Celestial Objects, Chinese Academy of Sciences, Kunming 650216, People's Republic of China\\
$^{3}$Center for Astronomical Mega-Science, Chinese Academy of Sciences, Beijing 100012, People's Republic of China \\
$^{4}$ University of Chinese Academy of Science, Beijing 100049, People's Republic of China  \\
$^{5}$ Kapteyn Astronomical Institute, University of Groningen, P.O. Box 800, NL-9700 AV Groningen, The Netherlands     \\
$^{6}$ Department of Physics, Xiangtan University, Xiangtan, Hunan 411105, People's Republic of China    \\
$^{7}$ Key Laboratory of Stars and Interstellar Medium, Xiangtan University, Xiangtan, Hunan 411105, People's Republic of China
}

\date{Accepted XXX. Received YYY; in original form ZZZ}

\pubyear{2019}

\begin{document}

\label{firstpage}
\pagerange{\pageref{firstpage}--\pageref{lastpage}}
\maketitle

\begin{abstract}
We have found and analysed 16 multi-peaked type-I bursts from the neutron-star low mass X-ray binary 4U 1636$-$53 with the Rossi X-ray Timing Explorer (RXTE). 
One of the bursts is a rare quadruple-peaked burst which was not previously reported.
All 16 bursts show a multi-peaked structure not only in the X-ray light curves but also in the bolometric light curves. 
Most of the multi-peaked bursts appear in observations during the transition from the hard to the soft state in the colour-colour diagram. 
We find 
an anti-correlation between the second peak flux and the separation time between two peaks.
We also find that in the double-peaked bursts the peak-flux ratio and the temperature of the thermal component in the pre-burst spectra are correlated. 
This indicates that the double-peaked structure in the light curve of the bursts may be affected by enhanced accretion rate in the disc, or increased temperature of the neutron star. 

\end{abstract}



\section{INTRODUCTION}
\label{INTRODUCTION}
Thermonuclear (Type I) X-ray bursts show a sudden increase in X-ray intensity, becoming $\sim 10-100$ times brighter than the persistent level, triggered by unstable ignition of accreted fuel on the surface of an accreting neutron star (NS) in low mass X-ray binaries (LMXBs) \citep{Galloway2008ApJS, Galloway2017arXiv}.
Type I X-ray bursts were first detected in 1975 in the binary $3A 1820-30$ in the globular cluster NGC 6624 \citep{Grindlay1976}; subsequently a growing population of bursters has been observed by different X-ray satellites \citep{Galloway2017arXiv}. 
In a typical X-ray burst, the light curve shows a single-peaked profile with a fast rise ($\sim 1-5$ s) and an exponential decay within $10-100$ s  \citep{Lewin1993SSRv,strohmayer_bildsten_2006,Galloway2008ApJS}.

Besides the single-peaked normal bursts, multi-peaked bursts have also been reported in previous studies. 
Double-peaked bursts have been reported in several NS-LMXBs, e.g.,  4U 1608$-$52 \citep{Penninx1989A&A}, GX 17+2 \citep{Kuulkers2002A&A}, 4U 1709$-$267 \citep{Jonker2004MNRAS} and MXB 1730$-$335 \citep{Bagnoli2014MNRAS}. 
With the Rossi X-ray Timing Explorer (RXTE), \cite{Watts2007AA} analyzed 4 double-peaked bursts in 4U 1636$-$53. 
Still more rare, bursts with triple-peaked structure have also been observed in 4U 1636$-$53 \citep{vanParadijs1986MNRAS, Zhanggb2009MNRAS}. 
While investigating the cooling phase of X-ray bursts in 4U 1636$-$53, \cite{Zhanggb11} reported $12$ double-peaked bursts and found that most of them appeared at the vertex of colour-colour diagram.
Recently, there were two new observations of double-peaked burst, one in the soft spectral state in 4U 1608$-$52 \citep{Jaisawal2019ApJ}, and another one in SAX J1808.4$-$3658 \citep{Bult2019ApJL}, both using the Neutron Star Interior Composition Explorer (NICER).

The double-peaked structures in the burst light curve can be separated into two groups. 
The first one consists of bursts with a double-peaked profile in X-rays but a single-peaked profile in the bolometric lightcurve, generally accompanied by photospheric radius expansion (PRE), where the flux of the burst reaches the Eddington luminosity. In this case, the temperature of the photosphere temporarily shifts out of the instrument passband, causing an apparent dip in the observed X-ray light curve \citep{Paczynski1983ApJ}. 
 The other group consists of bursts that have a double-peaked profile both in X-rays and the bolometric lightcurve. Most of these bursts have low peak flux, although PRE bursts with double-peaked profiles both in X-rays and bolometric luminosity have  been recently observed with NICER \citep{Jaisawal2019ApJ,Bult2019ApJL}.

Several theoretical models have been proposed to explain the double-peaked bursts.
\cite{Fujimoto88} proposed a model of stepped thermonuclear energy generation due to shear instabilities in the fuel  on the NS surface.  
\cite{Melia92} suggested that the double-peaked bursts are due to the scattering of the X-ray emission by material evaporated from the disk during the burst.   
These models, however, can not reproduce the observed double-peaked profiles in the light curve, black-body temperature and radius \citep{Bhattacharyya200601}. 
\cite{Fisker04} suggested that a waiting point impedes the nuclear reaction flow and causes a stepped release of thermonuclear energy, but this idea has difficulties in explaining the large dips observed between the two peaks \citep{Bhattacharyya200601}.  
The thermonuclear flame spreading model provided by \cite{Bhattacharyya200601} suggested that the double-peaked structure is caused by high latitude ignition and stalling approaching the equator. 
This model qualitatively explains the essential features of the light curve and reproduces the spectral evolution of two double-peaked bursts in 4U 1636$-$53.
However, one of the problems of the flame spreading model is that it can not explain the triple-peaked bursts \citep{Zhanggb2009MNRAS}.
\cite{Lampe2016} in their simulations found that low accretion rate and high metallicity could affect the burst morphology and produce twin-peaked structure when a large amount of hydrogen has been depleted.
Recently, \cite{Bult2019ApJL} suggested that the bright double-peaked bursts are due to the local Eddington limits associated with the hydrogen and helium layers of the NS envelope.
Understanding these mechanisms is important, because current models for multi-peaked X-ray bursts have met with only partial success in explaining their light curves and temperature profiles.

\begin{figure*}
\centering
\includegraphics[height=6.2in,width=6.0in, angle=0]{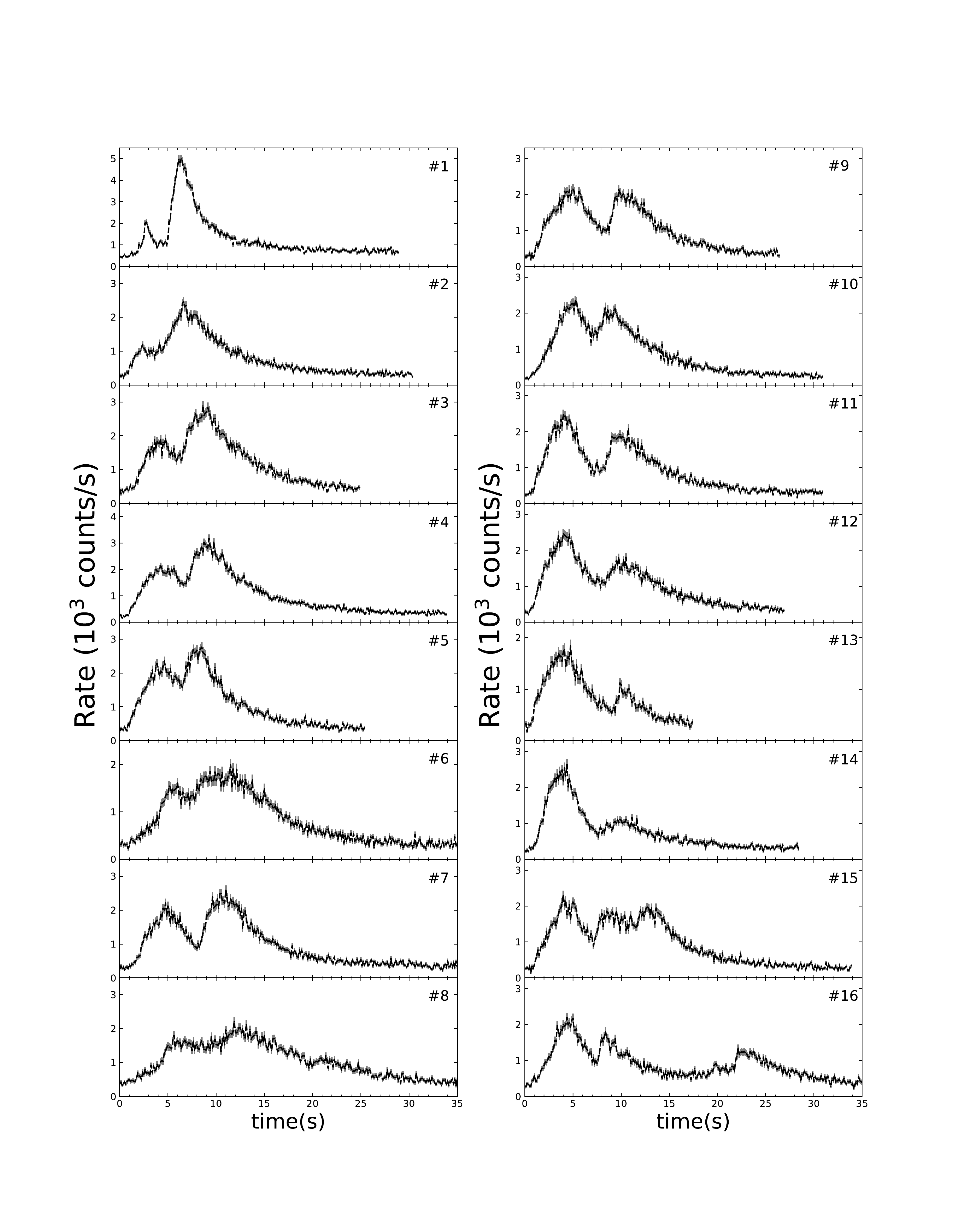} 
\caption{The light curves in the $2-60$ keV range of 16 multi-peaked bursts in 4U 1636$-$53 at 0.125s resolution observed with RXTE.
The 14 double-peaked bursts are defined as Class 1 bursts and are ordered by peak flux ratio (the ratio of the first peak flux to the second peak flux, $r_{1,2}$). 
Burst $\#15$ (three peaks) and burst $\#16$ (four peaks) are defined as Class 2 bursts.
}
\label{Figurex:the light curve}
\end{figure*}

Burst properties in an individual system depend mainly on accretion rate \citep{Fujimoto1981ApJ, Bildsten2000, Zhanggb11, Galloway2017arXiv}. 
For a specific source, given a certain global accretion rate, the local accretion rate varies with latitude, being higher at the equator and lower at high latitude \citep{Cooper07}. 
The ignition latitude depends on the column depth related to accretion rate to trigger a burst.
As the increasing global accretion rate, the lower latitude region firstly reach the critical local accretion rate and the fuel become stable burning, the ignition should occur at higher latitude even pole. 
Concerning the research of \cite{Cooper07} with the thermonuclear spreading model of \cite{Bhattacharyya200601},  \cite{Watts2007AA} expected to find more double-peaked bursts at higher global accretion rates than the single-peaked bursts.
However, \cite{Watts2007AA} presented the analysis of the accretion rate limited of 4 double-peaked bursts and posed a challenge to the above expectation.
In this paper, we collect a large sample to provide a more complete description of the observational features of multi-peaked bursts and further discuss the relation between accretion rate and multi-peaked structure.


In the time-resolved spectral analysis, the standard approach is to fit the X-ray burst spectra by assuming a constant persistent emission (non-burst component) during the burst \citep{Galloway2008ApJS}.
However, recent studies provide evidence of enhanced accretion during type-I X-ray bursts \citep{Worpel2013ApJ,Worpel2015ApJ}. 
The $f_a$ method gives improvements in the quality of spectral fit compared to the standard approach, and the analysis is sensitive to changes in the persistent spectrum in the $2.5-20$ keV \citep{Worpel2015ApJ}. 
In this paper, we adopt both the standard approach and the $f_a$ method to analyze our sample of multi-peaked bursts.

\begin{table*}
\caption{Properties of the light curves of Class 1 bursts in 4U 1636$-$53.
The column $n$ gives burst number sorted by the value of the peak flux ratio ($r_{1,2}$) which is the ratio of the first peak to the second peak flux.
We used a Gaussian function to fit each peak of every burst.
The quantify $t_{p,1}$ gives the peak time of the first peak, $\delta$ is the separation time between the first and the second peak, $t_{p,2}$ is the peak time of the second peak. 
}
\setlength{\tabcolsep}{9pt}
\begin{tabular}{l l l l l l l l}
\toprule   
&  Start time (UTC)& Obsid& $n$&  $r_{1,2}$&   $t_{p,1}$(s)&  $\delta$(s)& $t_{p,2}$(s)     \\
\hline
\multirow{14}*{Class1}& 2002-02-28 23:42:54.756& 60032-05-15-00&    $\#1$&  0.29$\pm$0.04& 1.6$\pm$0.3& 4.0$\pm$0.3& 5.6$\pm$0.1 \\   
\cline{2-8}
&2010-01-16 01:59:57.756& 95087-01-08-00& $\#2$& 0.40$\pm$0.08&  1.1$\pm$0.2&  5.1$\pm$0.3&  6.2$\pm$0.1 \\
\cline{2-8}
&2001-09-05 08:15:05.756&  60032-01-09-01& $\#3$& 0.61$\pm$0.08&  3.0$\pm$0.4& 4.9$\pm$0.4&  7.9$\pm$0.2\\
\cline{2-8}
&2009-02-20 03:43:55.756&  94087-01-29-00& $\#4$& 0.65$\pm$0.10&  3.1$\pm$0.2&   4.9$\pm$0.2&  8.0$\pm$0.1 \\
\cline{2-8}
&2001-10-03 00:22:20.756&  60032-01-13-01&  $\#5$& 0.73$\pm$0.09&  3.1$\pm$0.1&   4.0$\pm$0.1&  7.1$\pm$0.1   \\
\cline{2-8}
&2002-01-15 14:08:16.756&  60032-05-08-00&  $\#6$& 0.75$\pm$0.13&  4.2$\pm$0.2& 5.5$\pm$0.3& 9.8$\pm$0.2 \\
\cline{2-8}
&2006-02-06 16:21:20.756&  91024-01-72-10&  $\#$7& 0.79$\pm$0.11&  3.7$\pm$0.1&  6.3$\pm$0.2&  10.0$\pm$0.1\\
\cline{2-8}
&2008-05-20 01:15:35.256&  93087-01-66-10&  $\#$8& 0.81$\pm$0.20&   5.0$\pm$0.2&  6.6$\pm$0.3&  11.6$\pm$0.2\\
\cline{2-8}
&2002-01-08 12:22:46.256&  60032-01-19-000& $\#$9& 1.05$\pm$0.15&   3.6$\pm$0.1&  6.1$\pm$0.2&  9.6$\pm$0.2 \\
\cline{2-8}
&2005-10-09 06:01:21.756&  91024-01-12-10&  $\#$10& 1.10$\pm$0.15&   3.6$\pm$0.2&  4.7$\pm$0.3&  8.4$\pm$0.2\\
\cline{2-8}
&2008-11-12 01:18:29.256&  93087-01-56-20&  $\#$11& 1.34$\pm$0.23&   3.0$\pm$0.1&   6.4$\pm$0.1&   9.4$\pm$0.1\\
\cline{2-8}
&2006-08-08 16:40:14.256&  92023-01-80-00&  $\#$12& 1.56$\pm$0.22&   3.0$\pm$0.1&   6.3$\pm$0.1&   9.3$\pm$0.1\\
\cline{2-8}
&2009-12-21 03:44:05.756&  94087-01-83-10&  $\#$13& 2.24$\pm$0.62&   3.0$\pm$0.1&   6.4$\pm$0.3&   9.4$\pm$0.3 \\ 
\cline{2-8}
&2006-08-17 09:24:22.256&  92023-01-84-00&  $\#$14& 2.85$\pm$0.74&   2.7$\pm$0.1&  6.4$\pm$0.2&   9.1$\pm$0.2 \\
\bottomrule  

\end{tabular}
\vspace{1ex}

\label{Tablex:lc parameters of Class 1}
\end{table*}

The LMXB 4U 1636$-$53 is one of the best-studied sources of X-ray bursts. The NS is in a binary system in a 3.8 hr orbit \citep{Paradijs1990A&A} with an 18th magnitude blue star companion \citep{Galloway2008ApJS}, and the spin period of the NS is 581 Hz \citep{Strohmayer1998a, Strohmayer1998b}.
4U 1636$-$53 is an Atoll source, and as the source moves in the colour-colour diagram (hereafter CCD) from the top right to bottom right, the accretion rate gradually increases, with a transition from the Island to the Banana state \citep{Hasinger1989A&A}. 
The single peak bursts show a uniform distribution in the CCD \citep{Zhanggb11}.
About a dozen double-peaked and two triple-peaked X-ray bursts have been discovered from 
this source using different satellites \citep{Sztajno85, vanParadijs1986MNRAS, Lewin1987ApJ, Bhattacharyya200601, Bhattacharyya200604, Watts2007AA, Galloway2008ApJS, Zhanggb2009MNRAS, Zhanggb11}.  
The large multi-peaked bursts sample makes 4U 1636$-$53 an ideal source to study the properties and evolution of this kind of bursts.

\begin{table*}
\caption{Properties of the light curves of Class 2 bursts in 4U 1636$-$53.
The quantities $t_{p,3}$ and $t_{p,4}$ are the peak time of the third and fourth peak, respectively,
$\delta_{23}$ and $\delta_{34}$ are the separation time between the second and the third peak, as well as between the third and the fourth peak, respectively.
For the triple-peaked burst, we use the same method that we used in double-peaked bursts.
For the quadruple-peaked burst, we use two \textit{GAUSSIAN} functions adding two \textit{BURS} models to fit all data of the light curve (see detailed discussion in Sec \ref{The quadruple-peaked episode}).
}
\setlength{\tabcolsep}{6.7pt}
\begin{tabular}{l l l l l l l l l l l}
\toprule   
&Start time (UTC)& Obsid& $n$& $t_{p,1}$(s)&  $\delta$(s)& $t_{p,2}$(s)& $\delta_{23}$(s)& $t_{p,3}$(s)& $\delta_{34}$(s)& $t_{p,4}$(s) \\
\hline
\multirow{2}*{Class2}& 2006-12-11 18:23:26.756&  92023-01-44-10& $\#$15&  3.2$\pm$0.1&  4.4$\pm$0.1&  7.6$\pm$0.1&  4.0$\pm$0.2&  11.6$\pm$0.1&  $-$&  $-$\\
\cline{2-11}
&2010-12-26 20:26:19.756&  95087-01-82-10& $\#$16&  3.4$\pm$0.1&  3.7$\pm$0.1&  7.1$\pm$0.1&     11.8$\pm$0.2&  18.9$\pm$0.2&   2.7$\pm$0.2&  21.6$\pm$0.1 \\
\bottomrule  
\end{tabular}
\vspace{1ex}

          
\label{Tablex:lc parameters of Class 2}
\end{table*}

The structure of this paper is organised as follows. In Section \ref{OBSERVATIONS AND DATA ANALYSIS} we describe the data analysis of our sample. In Section \ref{RESULTS} we show the results of light curves and spectra. In Section \ref{DISCUSSION} we discuss our findings in the context of previous theoretical work.

\begin{figure*}
\centering
\includegraphics[height=10cm,width=13cm]{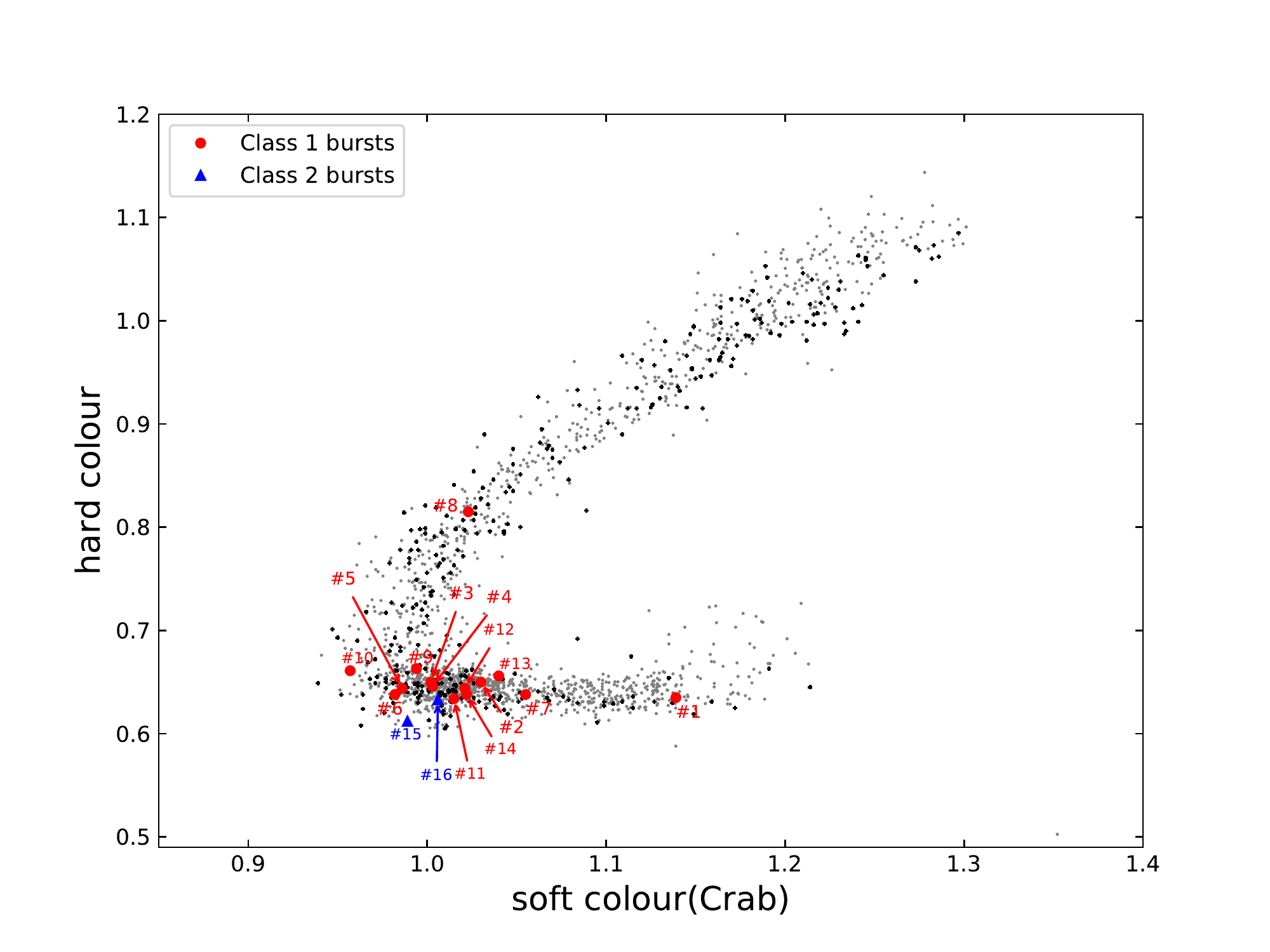}
\caption{Colour-colour diagram (CCD) of all RXTE observations of 4U 1636-53. 
The colours of 4U 1636$-$53 are normalized to the colours of Crab. 
Grey points represent all available observations with one point per observation. 
Black crosses stand for all X-ray burst in this source. 
Red filled circles and blue filled triangles represent Class 1 and Class 2 bursts, respectively.
}
\label{Figx:CCD}
\end{figure*}

\section{DATA ANALYSIS}
\label{OBSERVATIONS AND DATA ANALYSIS}

The Rossi X-ray Timing Explorer (RXTE) was launched in 1995, and operated until 2012 with a circular orbit at an altitude of $580$ km, correspoding to an orbital period of about $96$ min \citep{Bradt1993A+AS}.
We analysed all archived data from the proportional Counter Array (PCA) which is the main instrument onboard RXTE. The PCA consists of five collimated proportional counter units (PCUs), which are sensitive in the $2-60$ keV energy range with an energy resolution of $\sim 1$ keV at $6$ keV \citep{Jahoda2006ApJS}. 
For each observation we used the Standard2 data (16-s time resolution and 129 energy channels) to calculate X-ray colours.
We used the Standard1 mode (only the PCU2) to produce the burst light curves.
For the time-resolved spectral analysis of the bursts, we extracted  spectra in 64 channels from the Event data of all available PCUs.

We studied 336 type I X-ray bursts \citep[as in][]{Zhang2013MNRAS} in LMXB 4U 1636$-$53 with RXTE  and discover 16 multi-peaked bursts.
The 0.125s bin light curves of these 16 bursts in the $2-60$ keV are shown in  Figure \ref{Figurex:the light curve}. 
Following the procedure in \cite{Zhanggb11} and \cite{Zhang2013MNRAS}, we searched the $1$s Standard1 light curve for burst visually, and considered that the start time of a burst is when the flux is larger than 3 times the $1 \sigma$ error of the average persistent flux.
We find 14 double-peaked bursts \citep[4 bursts has investigated by] []{Watts2007AA}, one triple-peaked burst \citep{Zhanggb2009MNRAS} and one quadruple-peaked burst which was not reported in previous work.
We divided the 16 bursts into two classes according to the number of peaks. 
All the double-peaked bursts are classified as Class 1 bursts and the bursts with more than two peaks are classified as Class 2 bursts.

To study these bursts in detail, we introduce several parameters to characterise the burst light curve (see Table \ref{Tablex:lc parameters of Class 1} and Table \ref{Tablex:lc parameters of Class 2}). 
In Table \ref{Tablex:lc parameters of Class 1}, the column $n$ is the burst number sorted by the value of peak flux ratio, $r_{1,2}$ (see details later).
Because most of the data around the peak show a symmetric distribution, we used a Gaussian function to fit the data around each single peak in each burst light curve to get the peaking time.
The quantity $t_{p,1}$ is the peak time of the first peak, $\delta$ is the separation time between the first and second peak, and $t_{p,2}$ is the peak time of the second peak. 
In Table \ref{Tablex:lc parameters of Class 2}, we list the characteristics of the burst light curve in Class 2 bursts. 
The fitting process of triple-peaked burst is similar to that of the double-peaked bursts.
About the quadruple-peaked burst, we have a more detailed discussion in Sec \ref{The quadruple-peaked episode}.
The quantities $t_{p,3}$ and $t_{p,4}$ are the peak time of the third and fourth peak, respectively,
$\delta_{23}$ and $\delta_{34}$ are the separation time between the second and the third peak, as well as between the third and the fourth peak, respectively.
For all of these parameters we give the $1 \sigma$ error.

\begin{figure}
\centering
\includegraphics[height=4.5in,width=3.0in,angle=0]{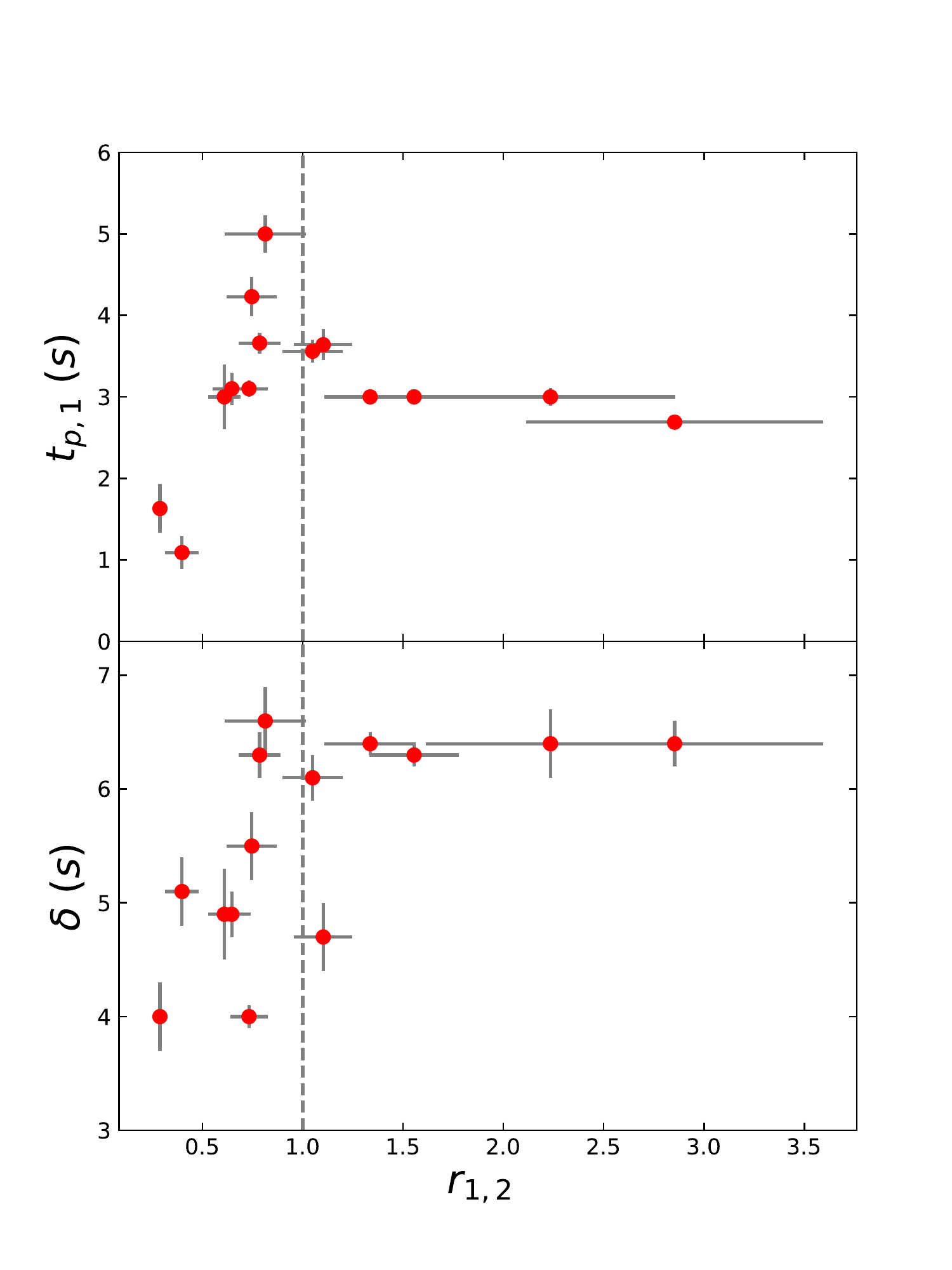}
\caption{
The peak time ($t_{p,1}$) of the first peak (upper panel) and the separation time ($\delta$) between the two peaks (lower panel) against the peak flux ratio ($r_{1,2}$) in the sample of the double-peaked bursts of 4U 1636$-$53.
Grey dashed line shows the ratio $r_{1,2}$ = 1. 
All these data are given in Table \ref{Tablex:lc parameters of Class 1}. 
}
\label{Figx:PD-PR}
\end{figure}

To trace the spectral state of the source when these multi-peaked bursts appear, we made a CCD as shown in Figure \ref{Figx:CCD}. 
We defined the soft colour as the ratio of the count rate in the $3.5-6.0$ keV to the count rate in the $2.0-3.5$ keV bands, and the hard colour as the ratio of the count rate in the $9.7-16.0$ keV to the count rate in the $6.0-9.7$ keV bands \citep{Zhanggb2009MNRAS}. 
The colours of the source are normalised by the Crab.  
For the colours of the source before the burst, we used $64$-s of the pre-burst spectrum. 
In Figure \ref{Figx:CCD}, the grey points represent all available observations. The black crosses represent all the bursts in this source. 
The red filled circles stand for Class 1 bursts, and the blue filled triangles indicate the Class 2 bursts.  
We find that most of the multiple-peaked bursts are located close to the vertex of the CCD.

We extracted the 64-s interval spectrum prior to a burst as persistent emission. We generated the instrument response matrix using the tool \textit{pcarsp} and the instrumental background using the tool \textit{pcabackest} in HEAsoft for each spectrum.  
In this work, we fitted the spectrum in the $3.0-20$ keV band using XSPEC version $12.10.1$ \citep{Arnaud1996ASPC}.
We added a $0.5 \%$ systematic error to the pre-burst spectra because of calibration uncertainties.
During the fitting process, we included the effect of interstellar absorption using the cross-sections of \cite{Balucinska-Church1992ApJ} and solar abundances from \cite{Anders1989}, with a fixed hydrogen column density, $N_{\rm H}$, of $0.36 \times 10^{22} \rm \, cm^{-2}$ \citep{Pandel2008ApJ}.

\newpage
\begin{table*}
\caption{Best-fitting parameters of the persistent emission before the 16 multi-peaked bursts in 4U 1636$-$53 with the model TBabs*(bbodyrad+powerlaw).
The quantity $kT_{\rm bb}$ is the blackbody temperature, $K_{\rm bb}$ is the normalisation of the blackbody, 
$\alpha$ is the power-law index, $N$ is the normalisation of the power law,  
$F_{\rm p}$ represents the persistent unabsorbed flux in the $2.5-25$ keV band and $\gamma$ is the luminosity of the source before the burst in Eddington units.
The spectral parameters of our best-fitting model are given with 1 $\sigma$ error.}
\setlength{\tabcolsep}{8pt}
\begin{tabular}{c c c c c c c c c c}
\toprule   
 & Obsid& $n$&  $\chi^{2}_{\nu}$&   $kT_{\rm bb}$ &    $K_{\rm bb}$&           $\alpha$&    $N$&  $F_{\rm p}$&  $\gamma$\\
 &  &  &  &  (keV)&  &  &  &  $(10^{-9} \rm erg s^{-1} cm^{-2})$&  $L_{\rm p}/L_{\rm Edd}$ \\
\hline
\multirow{14}*{Class1}& 60032-05-15-00& $\#$1&  1.19& 1.68$\pm$0.03& 27.90$\pm$1.72&  2.72$\pm$0.04& 3.38$\pm$0.25&  5.31& 13.0$\%$ \\   
\cline{2-10}
&95087-01-08-00& $\#$2&  1.89&  1.78$\pm$0.06&    8.10$\pm$0.98& 2.71$\pm$0.09& 1.44$\pm$0.19&  2.17& 5.3$\%$ \\
\cline{2-10}
&60032-01-09-01&  $\#$3&    1.75&  1.79$\pm$0.04&   10.64$\pm$0.74&  2.92$\pm$0.07&    2.60$\pm$0.27&  2.81& 6.9$\%$ \\
\cline{2-10}
&94087-01-29-00&  $\#$4&    0.90&  1.78$\pm$0.06&   7.68$\pm$0.94&    2.84$\pm$0.09&    1.95$\pm$0.25&  2.24&  5.5$\%$ \\
\cline{2-10}
&60032-01-13-01&  $\#$5&  1.08&  1.84$\pm$0.04&   8.81$\pm$0.66&    3.03$\pm$0.11&    2.63$\pm$0.38&  2.44&  6.0$\%$  \\
\cline{2-10}
&60032-05-08-00&  $\#$6&   1.42&  1.75$\pm$0.06&   7.26$\pm$1.02&    2.62$\pm$0.08&    1.28$\pm$0.15&  2.09&  5.1$\%$  \\
\cline{2-10}
&91024-01-72-10&  $\#$7&    0.82&  1.69$\pm$0.05&   11.92$\pm$1.26&   2.77$\pm$0.07&    1.96$\pm$0.22&  2.61&  6.4$\%$ \\
\cline{2-10}
&93087-01-66-10&  $\#$8&    2.20&  1.57$\pm$0.12&   10.20$\pm$3.99&   2.20$\pm$0.09&    0.81$\pm$0.16&  2.60&   6.4$\%$ \\
\cline{2-10}
&60032-01-19-000& $\#$9&  1.00&  1.82$\pm$0.06&   6.66$\pm$0.75&    2.78$\pm$0.09&    1.75$\pm$0.22&   2.20&  5.4$\%$  \\
\cline{2-10}
&91024-01-12-10&  $\#$10&    1.22&  1.79$\pm$0.07&   5.15$\pm$0.74&    2.71$\pm$0.11&    1.11$\pm$0.17&   1.58&  3.9$\%$  \\
\cline{2-10}
&93087-01-56-20&  $\#$11&   1.09&  1.79$\pm$0.06&   8.70$\pm$0.96&    2.92$\pm$0.13&    1.90$\pm$0.35&   2.15&  5.3$\%$ \\
\cline{2-10}
&92023-01-80-00&  $\#$12&    0.92&  1.81$\pm$0.05&   8.10$\pm$0.79&    2.81$\pm$0.09&    1.99$\pm$0.24&   2.44&  6.0$\%$  \\
\cline{2-10}
&94087-01-83-10&  $\#$13&  1.54&  1.87$\pm$0.05&   8.33$\pm$0.07&    3.06$\pm$0.15&    2.66$\pm$0.51&   2.40&  5.9$\%$ \\
\cline{2-10}
&92023-01-84-00&  $\#$14&  2.25&  1.87$\pm$0.09&   6.73$\pm$1.07&    2.93$\pm$0.18&    2.12$\pm$0.53&   2.19&  5.4$\%$ \\
\bottomrule  
\multirow{2}*{Class2}&92023-01-44-10& $\#$15&  1.11& 1.86$\pm$0.04& 5.93$\pm$0.46& 3.07$\pm$0.13& 1.94$\pm$0.31&  1.70&  4.2$\%$ \\
\cline{2-10}
& 95087-01-82-10& $\#$16& 1.38&  1.82$\pm$0.06&  10.02$\pm$1.15& 3.18$\pm$0.10& 2.92$\pm$0.98&   2.34&  5.7$\%$    \\
\bottomrule  
\end{tabular}
\vspace{1ex}

\label{Tablex:pre-b fitted par 16 bb}                 
\end{table*}

\newpage
\begin{table*}
\caption{Best-fitting parameters of the persistent emission before the 16 multi-peaked bursts in 4U 1636$-$53 with the model TBabs*(diskbb+powerlaw).
The quantity $kT_{\rm dbb}$ is the disc blackbody temperature, $K_{\rm dbb}$ is the normalisation of the disk blackbody.
$\alpha$ is the power-law index, $N$ is the normalisation of the power law,  
$F_{\rm p}$ represents the persistent unabsorbed flux in the $2.5-25$ keV band and $\gamma$ is the luminosity of the source before the burst in Eddington units.
The spectral parameters of our best-fitting model are given with 1 $\sigma$ error.}
\setlength{\tabcolsep}{7.6pt}
\begin{tabular}{c c c c c c c c c c c c}
\toprule   
 & Obsid&        $n$&  $\chi^{2}_{\nu}$&   $kT_{\rm dbb}$&  $K_{\rm dbb}$&  $\alpha$&  $N$&  $F_{\rm p}$&  $\gamma$ \\
&  &  &  &  (keV)&  &  &  &  $(10^{-9} \rm erg s^{-1} cm^{-2})$&  $L_{\rm p}/L_{\rm Edd}$\\
\hline
\multirow{14}*{Class1}& 60032-05-15-00& $\#$1&  0.72& 2.42$\pm$0.05& 6.60$\pm$0.59&  2.62$\pm$0.12&  1.48$\pm$0.29&   5.20&  12.8$\%$ \\   
\cline{2-10}
&95087-01-08-00&  $\#$2&   2.02&   2.66$\pm$0.13&   1.68$\pm$0.26&    2.80$\pm$0.31&    0.97$\pm$0.38&  2.14&  5.3$\%$ \\
\cline{2-10}
&60032-01-09-01&  $\#$3&   1.90&    2.66$\pm$0.05&   2.38$\pm$0.16&   3.32$\pm$0.37&    2.46$\pm$1.10&  2.76&  6.8$\%$ \\
\cline{2-10}
&94087-01-29-00&  $\#$4&   0.84&   2.65$\pm$0.11&   1.64$\pm$0.22&    3.02$\pm$0.29&    1.61$\pm$0.55&  2.21&   5.4$\%$\\
\cline{2-10}
&60032-01-13-01&  $\#$5&   1.13&   2.68$\pm$0.04&   2.19$\pm$0.26&    3.94$\pm$0.63&    4.17$\pm$3.42&  2.41&  5.9$\%$\\
\cline{2-10}
&60032-05-08-00&  $\#$6&   1.37&   2.59$\pm$0.14&   1.56$\pm$0.31&    2.62$\pm$0.19&    0.83$\pm$0.22&  2.05&  5.0$\%$\\
\cline{2-10}
&91024-01-72-10&  $\#$7&   0.72&   2.47$\pm$0.10&   2.68$\pm$0.38&   2.81$\pm$0.20&    1.25$\pm$0.33&  2.56&  6.3$\%$\\
\cline{2-10}
&93087-01-66-10&  $\#$8&   2.22&   2.31$\pm$0.24&   2.35$\pm$1.24&    2.05$\pm$0.20&    0.47$\pm$0.21&  2.58&  6.3$\%$\\
\cline{2-10}
&60032-01-19-000& $\#$9&   1.27&   2.74$\pm$0.12&   1.38$\pm$0.18&    2.96$\pm$0.34&    1.45$\pm$0.63&  2.16&   5.3$\%$\\
\cline{2-10}
&91024-01-12-10&  $\#$10&  1.12&    2.70$\pm$0.14&   1.07$\pm$0.18&    2.86$\pm$0.32&    0.84$\pm$0.33&  1.54&  3.8$\%$\\
\cline{2-10}
&93087-01-56-20&  $\#$11&  0.98&   2.63$\pm$0.08&   1.99$\pm$0.22&    3.35$\pm$0.61&    1.67$\pm$1.45&  2.10&  5.2$\%$\\
\cline{2-10}
&92023-01-80-00&  $\#$12&  0.90&   2.72$\pm$0.08&   1.72$\pm$0.16&    3.10$\pm$0.34&    1.75$\pm$0.70&  2.40&  5.9$\%$\\
\cline{2-10}
&94087-01-83-10&  $\#$13&  1.41&   2.71$\pm$0.05&   2.10$\pm$0.30&    4.04$\pm$0.75&    4.41$\pm$4.76&   2.37&  5.8$\%$\\
\cline{2-10}
&92023-01-84-00&  $\#$14&  2.17&   2.78$\pm$0.09&   1.58$\pm$0.35&    3.66$\pm$0.94&    2.94$\pm$4.85&  2.16&  5.3$\%$\\
\bottomrule  
\multirow{2}*{Class2}&92023-01-44-10& $\#$15&  1.14& 2.69$\pm$0.05& 1.53$\pm$0.19& 4.13$\pm$0.62& 3.62$\pm$2.85&   1.68&  4.1$\%$\\
\cline{2-10}
& 95087-01-82-10& $\#$16& 1.45&  2.53$\pm$0.07&  3.03$\pm$0.46& 5.80$\pm$1.59& 29.94$\pm$108.07&   2.37&  5.8$\%$\\
\bottomrule  

\end{tabular}
\vspace{1ex}

\label{Tablex:pre-b fitted par 16 diskbb}                  
\end{table*}

We used all available PCUs during the X-ray bursts to produce time-resolved spectrum.
We corrected every spectrum for dead time according to the methods supplied by the RXTE team.
Since the light curve of bursts decay is quite smooth, to compensate the lower count rates, we extracted spectrum over longer intervals in the tail of the bursts.
For each multiple-peaked burst, we generated one instrument response matrix using the \textit{pcarsp} and the instrumental background using the \textit{pcabackest} in HEAsoft.

For the burst spectral analysis, we initially used a single-temperature blackbody model, TBabs*bbodyrad, to fit the net burst spectra, which is well established as a standard procedure in X-ray burst analysis \citep{Kuulkers2002A&A, Galloway2008ApJS}.
The model provides the blackbody colour temperature, $kT_{\rm bb}$, and the normalization, $K_{\rm bb}$, proportional to the square of the blackbody radius of the burst emission surface, and it allows us to estimate the bolometric luminosity as a function of time assuming a distance of $5.95$ kpc \citep{Fiocchi2006ApJ}.  
The bursts bolometric flux are calculated as
\begin{equation}
    F=1.076\times10^{-11}\left(\dfrac{kT_{\rm{bb}}}{1\,\rm{keV}}\right)^{4}K_{\rm{bb}}\quad \rm{ergs\;cm^{-2}\;s^{-1}},
\end{equation}
where $K_{\rm bb} = R^{2}_{\rm km}/D^{2}_{10}$,
$R_{\rm km}$ is the effective radius of the emitting surface in km, and $D_{10}$ is distance to the source in units of 10 kpc.
We defined $r_{1,2}$ as the ratio of the first peak to the second peak flux.
We note that, to reduce the effect of the instrument, we used the bolometric flux (not the net burst count rate ratio in Figure \ref{Figurex:the light curve}) from the standard approach to calculate the ratio $r_{1,2}$.

After subtracting the instrumental background for the time-resolved spectra, 
we used another model that allows us to vary the pre-burst spectrum by a free scaling factor \citep{Worpel2013ApJ}.
We used two models to describe the persistent emission: TBabs*(bbodyrad + powerlaw) and TBabs*(diskbb + powerlaw) in XSPEC.  
The fit results for the above two models are shown in Tables \ref{Tablex:pre-b fitted par 16 bb} and \ref{Tablex:pre-b fitted par 16 diskbb}, respectively.  
In Table \ref{Tablex:pre-b fitted par 16 bb}, $kT_{\rm bb}$ is the blackbody temperature and $K_{\rm bb}$ is the normalization of the balckbody, 
$\alpha$ is the power-law index and $N$ is the normalization of the power law.
The spectral parameters of our best-fitting model are given with 1 $\sigma$ error.
In Table \ref{Tablex:pre-b fitted par 16 diskbb}, $kT_{\rm dbb}$ is the disk blackbody temperature and $K_{\rm dbb}$ is the normalization of the disk balckbody.
Comparing the fitting results of the two pre-burst spectra models (see Table \ref{Tablex:pre-b fitted par 16 bb} and Table \ref{Tablex:pre-b fitted par 16 diskbb}), we selected the TBabs*(bbodyrad + powerlaw) as the best pre-burst model,
so we used TBabs*($f_{a}$*(bbodyrad + powerlaw) + bbodyrad) in our analysis (the so-called $f_{a}$ method) to re-fit the net burst spectra.
The parameter $f_{a}$ was allowed to vary between $-100$ and $100$ during the fits.

In Tables \ref{Tablex:pre-b fitted par 16 bb} and \ref{Tablex:pre-b fitted par 16 diskbb}, $F_{\rm p}$ represents the persistent unabsorbed flux in the $2.5-25$ keV band and $\gamma$ is the luminosity of the source before the burst in Eddington units.
We used the unabsorbed $2.5-25$ keV flux, $F_{\rm p}$, a bolometric correction factor $c_{\rm bol}=1.2$ \citep{Galloway2008ApJS}, and the source distance of $5.95 \, \rm kpc$ to estimate the X-ray persistent luminosity, $L_{\rm p}$. 
In the calculation of Eddington luminosity we assume that the ratio of the colour temperature to the effective temperature, $T_{c}/T_{e}$, is 1.4 \citep{Madej2004ApJ}, the NS mass is $1.4 \, M_{\odot}$ and the hydrogen mass fraction $\rm X=0.7$. 
We note that, in the model of TBabs*(bbodyrad+powerlaw) $T_{c}$ corresponds to $kT_{\rm bb}$ and in the model of TBabs*(diskbb+powerlaw) $T_{c}$ corresponds to $kT_{\rm dbb}$.

\section[]{RESULTS}
\label{RESULTS}
\subsection{Multi-peaked bursts light curves}
\label{Multi-peaked bursts light curves}
Figure \ref{Figurex:the light curve} shows the 14 bursts with  double-peaked profiles.
The bursts display large or gentle dips in their light curves. The first peak can be either weaker/shorter or stronger/longer than the second peak.
In Figure \ref{Figx:PD-PR} we show the rise time, $t_{p,1}$, of the first peak and the separation time, $\delta$, of the two peaks as a function of the peak-flux ratio, $r_{1,2}$.
The grey dashed line represents $r_{1,2}$ = 1. 
The rise time of the first peak is in the range of $\sim1-6$s, and the separation in time between the two peaks is in the range of $\sim3-7$s.
We find that: (1) in the case of $r_{1,2}$ $< 1$, as the peak flux ratio increases, 
both the rise time of the first peak and the separation in time increase;
(2) in the case of $r_{1,2}$ $> 1$, the above parameters do not depend upon peak flux ratio.

Besides the triple-peaked burst (burst $\#15$) reported by \cite{Zhanggb2009MNRAS}, we also discovered a $\sim40 \, \rm s$ long quadruple-peaked burst (burst $\#16$; date of observation: 2006 August 17), which was not reported in previous studies.

\subsection{Multi-peaked bursts in the CCD}
\label{Multi-peaked bursts on the CCD}

Figure \ref{Figx:CCD} shows the position of all multi-peaked bursts in the CCD of 4U 1636$-$53. The normal, single-peaked, bursts observed with RXTE are distributed more or less uniformly across the CCD. 
However, except for burst $\#1$ and $\#8$, most of the multiple-peaked bursts are located close to the vertex of the CCD.

With the lowest peak flux ratio ($r_{1,2}\sim0.3$) and highest peak flux ($\sim5.2 \times 10^{-8}\; \rm ergs \; cm^{-2} \; s^{-1} $ using the standard approach), burst $\#1$ is located at the bottom right, when the source was in the so-called upper banana branch in the CCD.  
In the light curve of burst $\#1$, the first peak is relative weak ($r_{1,2} \sim 0.3$) and the whole burst is dominated by the second peak.
Burst $\#1$ has also the second shortest rise time ($t_{p,1} \sim 1.6 \, \rm s$) of the first peak and shortest separation time ($\delta$ $\sim4.0 \, \rm s$).

Burst $\#8$ is located at the position between the Banana and the Island states in the CCD. 
Different from burst $\#1$, burst $\#8$ has the longest rise time ($t_{r,1} \sim 5.0 \, \rm s$) of the first peak and the longest separation time ($\delta$ $\sim6.6 \, \rm s$), and a high $r_{1,2}$ ($\sim 0.8$).

\subsection{The relation between pre-burst spectrum and burst profile}
\label{The relation between pre-burst spectrum and burst profile}

\begin{figure*}
\centering
\includegraphics[height=8cm,width=8cm]{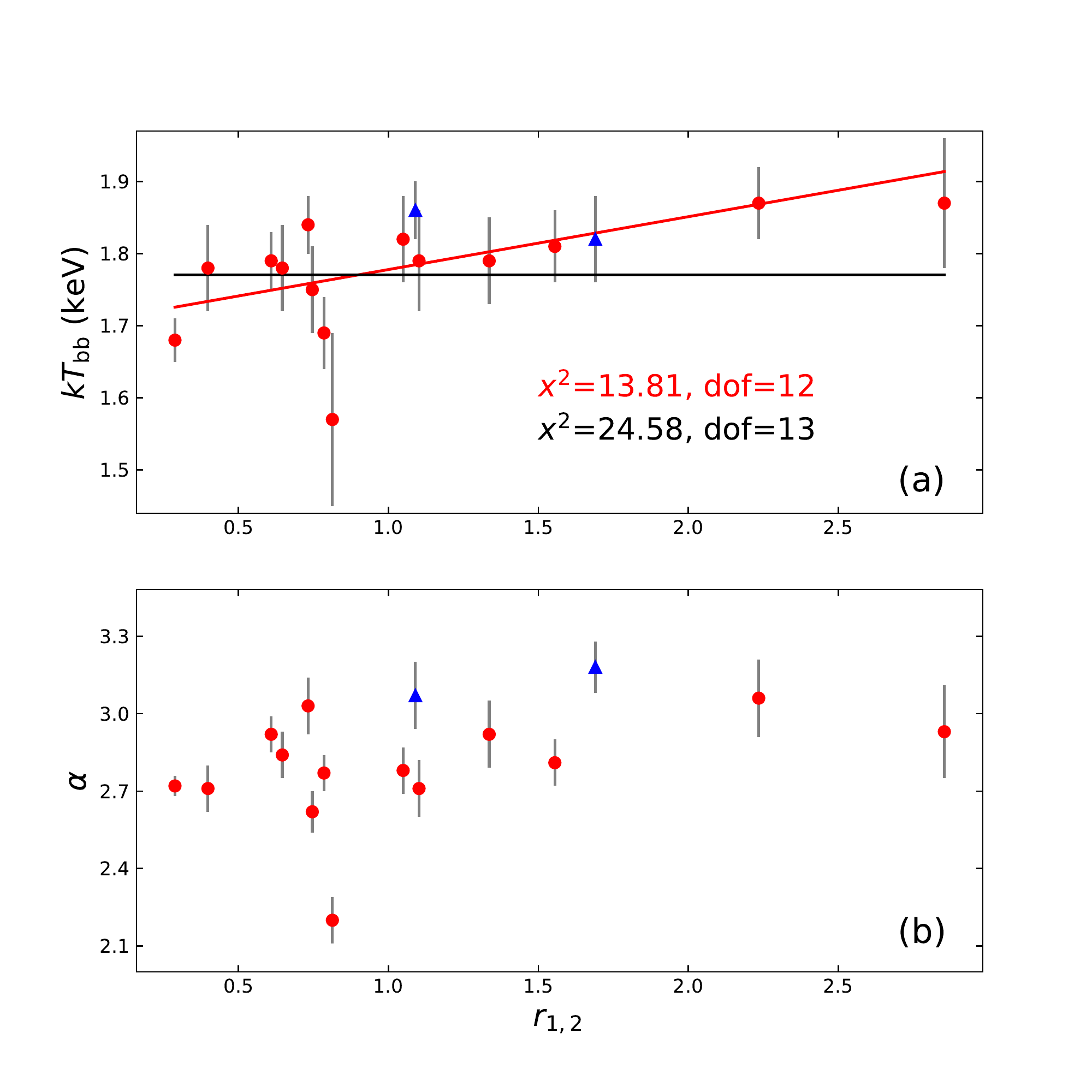}
\includegraphics[height=8cm,width=8cm]{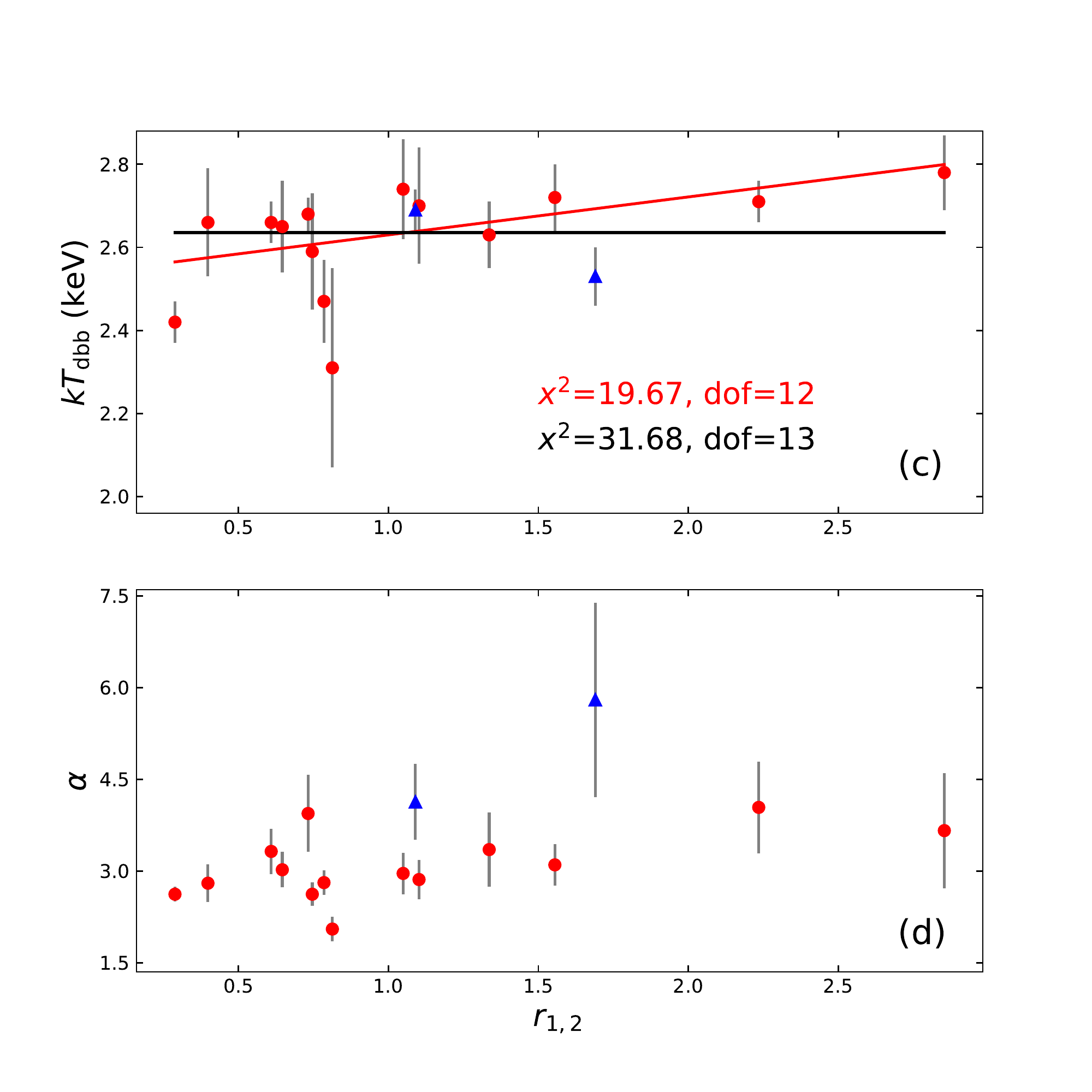}
\caption{Parameter of the persistent spectrum before the multi-peaked burst in 4U 1636$-$53. Left panels: blackbody temperature ($kT_{\rm bb}$), and power-law index ($\alpha$) for the model TBabs*(bbodyrad+powerlaw) against the peak flux ratio ($r_{1,2}$);
right panels: the disk blackbody temperature ($kT_{\rm dbb}$) and the power-law index ($\alpha$) for the model TBabs*(diskbb+powerlaw) against the peak flux ($r_{1,2}$). 
Red dots and blue triangles represent Class1 and Class 2 bursts, respectively. 
We use a linear (red line) and constant (black line) function to fit only the Class 1 bursts data. 
}
\label{Figx:pre-par_PR}
\end{figure*}

To investigate the relation between the pre-burst spectra and the multi-peaked burst light curve profiles, we compared the spectral parameters of the persistent emission of Class 1 bursts with two properties of the light curve: the peak flux ratio, $r_{1,2}$, and the separation time of two peaks, $\delta$.
Figure \ref{Figx:pre-par_PR} shows the spectral parameters of the persistent emission (in two models) against the peak flux ratio.  
The red dots represent the Class 1 bursts, and the blue triangles represent Class 2 bursts. 
The left panels (a) and (b) in Figure \ref{Figx:pre-par_PR} display, respectively, the blackbody temperature, $kT_{\rm bb}$, and power-law index, $\alpha$, against the peak flux ratio, $r_{1,2}$, for the model TBabs*(bbodyrad+powerlaw). 
The blackbody temperature ranges from $1.57$ keV to $1.87$ keV. 
There appears to be a positive correlation between the blackbody temperature and the peak flux ratio. 
In order to check that,  we firstly fit these data with a constant model and get $\chi^2 = 24.6$ for 13 d.o.f.
We then fit these data with a line function, and get a slope of $0.07 \, \pm \, 0.02$ with a $\chi^2 = 13.8$ for 12 d.o.f. 
The F-test probability for these two fits is $0.009$, indicating that a linear function is slightly better than a constant.

The right panels (c) and (d) in Figure 4 show, respectively, the disk-blackbody temperature, $kT_{\rm dbb}$, and power-law index, $\alpha$, against the peak flux ratio, $r_{1,2}$ for the model TBabs*(diskbb+powerlaw). 
For panel (c) of Figure \ref{Figx:pre-par_PR}, we do the same analysis as in panel (a), obtaining a $\chi^2=31.7$ for 13 d.o.f. for a constant model and a slope of $0.09 \, \pm \, 0.03$ with a reduced chi-square of 1.64 ($\chi^{2}/dof$ = 19.7/12) for the line function. 
The F-test probability for these two fits is $0.019$. 
The above analysis indicates that the peak flux ratio, $r_{1,2}$, is marginally correlated with the temperature of the thermal component in the pre-burst spectra.

\begin{figure}
    \centering
        \includegraphics[height=6.4in,width=3.6in,angle=0]{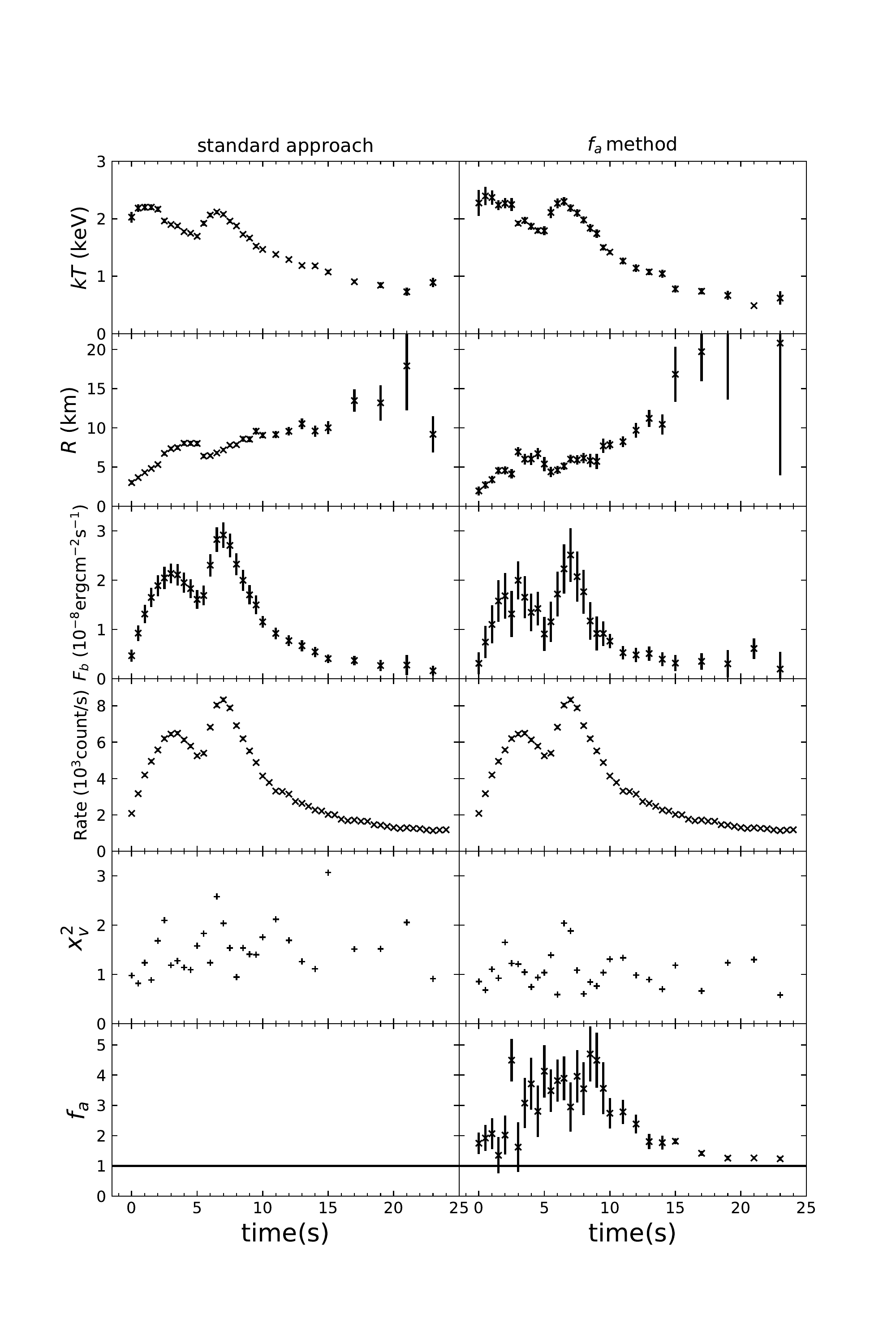}
    \caption{An example from Obsid 60032-01-13-01 of the spectral evolution of double-peaked burst spectra in 4U 1636$-$53 using two burst models. 
    Left panel: standard approach, right panel: $f_{a}$ method. 
    From top to bottom we display the blackbody temperature, the radius of the emitting surface, the bolometric flux, the $2-60$ keV light curve at 0.5s bin extracted from all available PCUs, $\chi^{2}_{\nu}$, and $f_{a}$.
    The horizontal solid line in the bottom pannel represents $f_{a}$=1.
    }
    \label{Fig: Obsid60032-01-13-01}
\end{figure}

\subsection{Time-resolved burst spectra}
\label{double-peaked bursts}
We adopted both the standard approach and the $f_{a}$ method to fit the time-resolved burst spectra. 
An example (burst $\#5$) of the best-fitting parameters using standard approach is shown in the left panel of Figure \ref{Fig: Obsid60032-01-13-01}. 
From the top to the third panel, we display the blackbody temperature, blackbody radius and bolometric flux as a function of time. 
We used a 0.5-s bin light curve in the $2-60$ keV range in the fourth panel. 
The double-peaked structures occur simultaneously in the X-ray and bolometric light curves, which indicates that the double-peaked profiles during the X-ray bursts is not due to a passband effect of the instrument. 
The time-resolved temperature shows two local maxima, as is also the case in the bolometric flux curve, however, the peaking time in temperature is different from the peaking time in bolometric flux curve.
After initially growing, the radius continues increasing following a dip,
and then remains more or less constant. 
The double-peaked structures are also apparent in the bolometric flux in the rest of the bursts of Class 1 bursts.

\begin{figure}
\includegraphics[height=6.0cm,width=8.0cm]{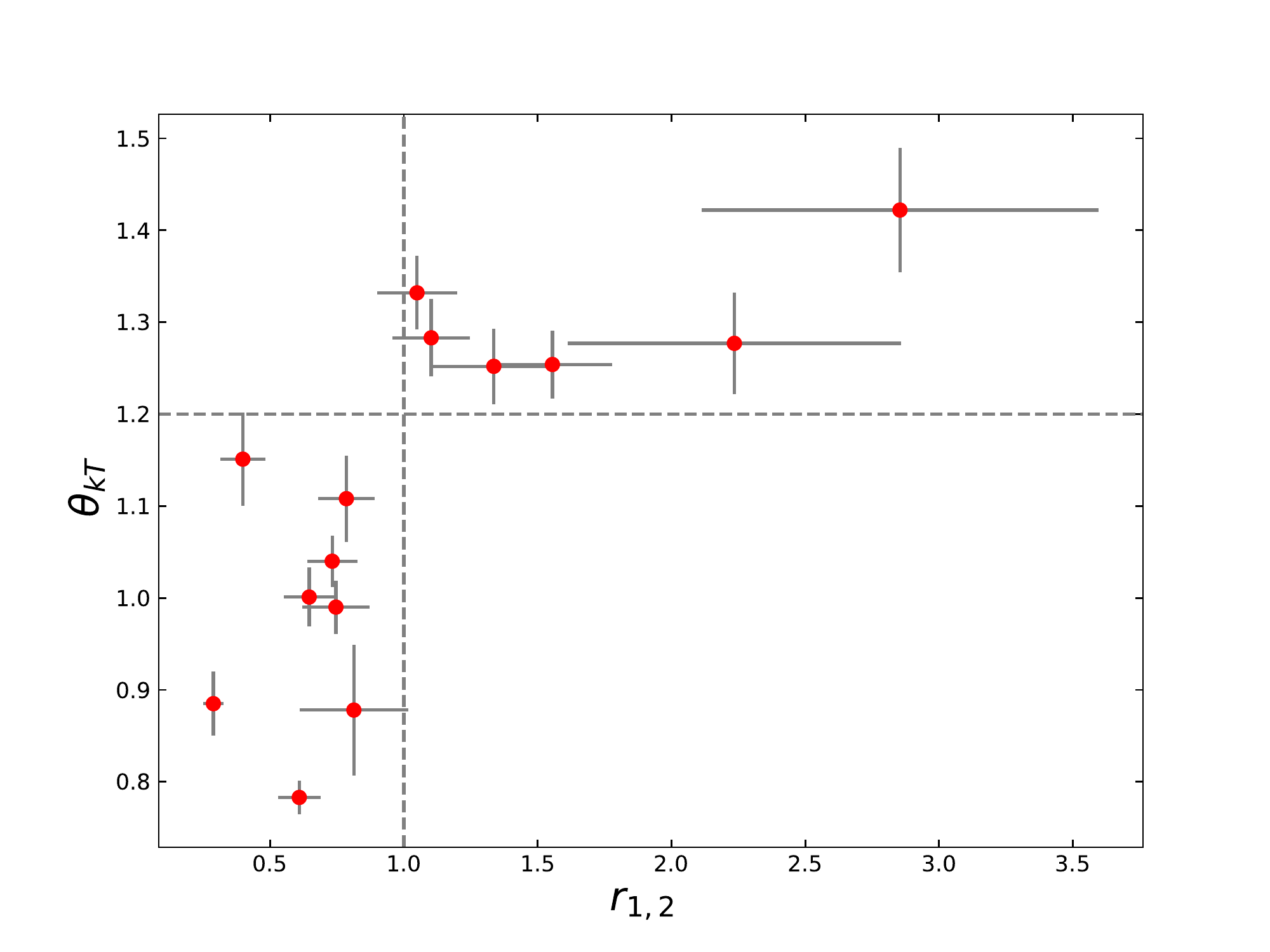}
\caption{The peak temperature ratio ($\theta_{kT}$) against the peak flux ratio ($r_{1,2}$) for the double-peaked bursts in 4U 1636$-$53 obtained from fits using the standard procedure to fit time-resolved spectra of X-ray bursts. 
The vertical and horizontal dashed lines correspond to the $r_{1,2}$=1.0, $\theta_{kT}$=1.2 respectively.
We note that the local maxima of temperature and flux do not necessarily occur at the same time.
}
\label{Fig:TPR_PR}
\end{figure}

We then measured the local peak temperature of the double-peaked bursts in the standard approach. The peak temperature ratio, $\theta_{kT}$ represents the ratio of the first peak to the second peak temperature.
Figure \ref{Fig:TPR_PR} shows the peak temperature ratios against the peak flux ratios in the 14 double-peaked bursts. 
The vertical and horizontal dashed lines represent $r_{1,2}$=1.0 and $\theta_{kT}$=1.2, respectively.
When $r_{1,2} > 1$, the peak temperature ratios are always larger than 1.2. In the case of $r_{1,2} < 1$, the first peak temperature can be higher or lower than the second peak temperature and $\theta_{kT}$ is always less than 1.2. 
There appears to be a bimodal distribution in the temperature ratios.

\begin{figure}
    \centering
        \includegraphics[height=4.4in,width=3.3in,angle=0]{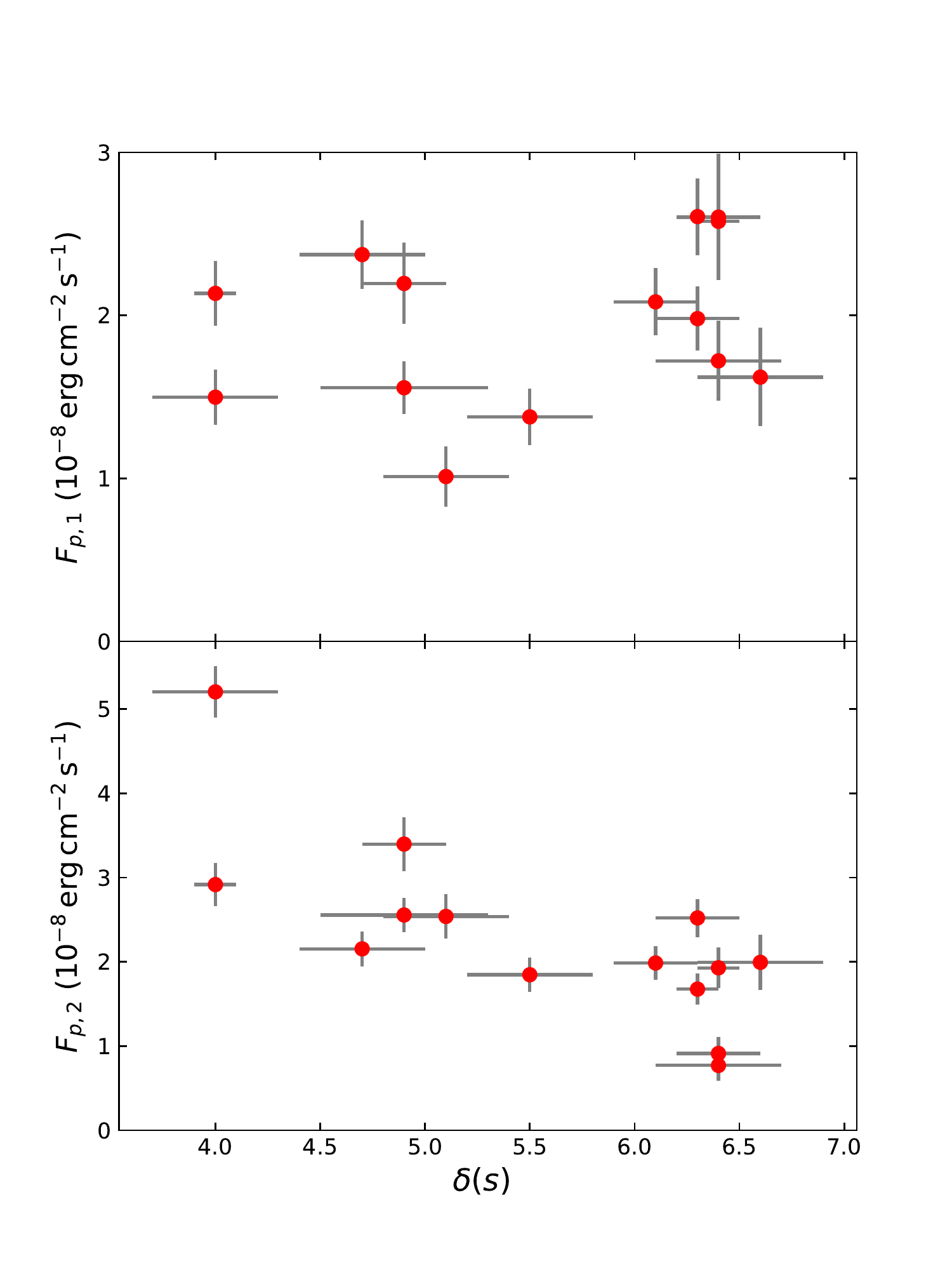}
    \caption{The first ($F_{p,1}$) and second ($F_{p,2}$) peak flux using the standard approach against the separation time ($\delta$) between the two peaks of 14 double-peaked bursts in 4U 1636$-$53.
    }
    \label{Fig:Flux1-Flux2_duration}
\end{figure}

Figure \ref{Fig:Flux1-Flux2_duration} shows the first and second peak flux in the standard approach against the duration time between two peaks, respectively. 
The typical PRE peak flux is in the range of $(6-8) \times 10^{-8} \; \rm ergs \; \rm cm^{-2} \; \rm s^{-1}$ in 4U 1636$-$53 \citep{Lyu2015MNRAS}. In our present sample, there are no PRE events.
There is no significant trend between the first peak flux and the duration time, however, an anti-correlation is present between the second peak flux and the separation time between two peaks.

We show the fits results of burst $\#5$ with the $f_{a}$ method in the right-hand panel of Figure \ref{Fig: Obsid60032-01-13-01}. As in the standard approach, both the bolometric flux and the X-ray light curve show a double-peaked structure. 
The evolution of black body temperature and radius in the $f_{a}$ method is similar to that in the standard approach.
The increased black body radius in the cooling phase may be explained in two ways. One would be the influence of persistent emission (accretion geometry) in the soft state \citep{Kajava2014MNRAS}, the other would be different canonical composition in the NS atmosphere \citep{Suleimanov2011ApJ, Zhanggb11}. 
The $f_{a}$ plotted in the bottom panel shows values larger than one during the whole burst.
Due to the new parameter $f_{a}$, the bolometric flux is lower than in the standard approach and has larger error bars.  
The horizontal solid line in the bottom panels stands for $f_{a}$=1. 
In general, the reduced $\chi^2$ distribution in the $f_{a}$ method is lower than that in the standard approach.

\begin{figure}
    \centering
        \includegraphics[height=8.0in,width=3.6in]{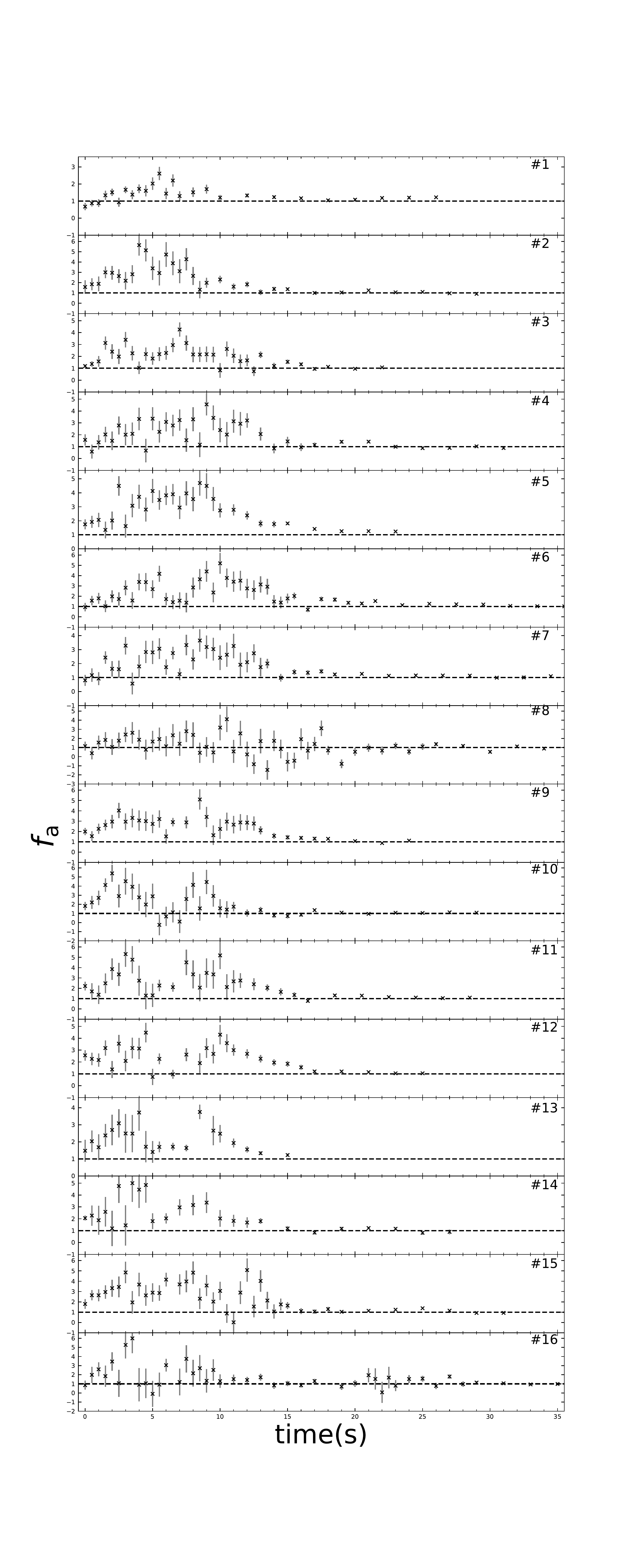}
    \caption{The time evolution of the $f_{a}$ factor of the 16 multi-peaked bursts in the 4U 1636$-$53. 
    Black horizontal dashed lines represent $f_{a}=1$. 
    The sequence of the bursts is the same as in Figure \ref{Figurex:the light curve}.
    The value of $f_{a}$ changes with the bolometric flux and is larger than 1 during the whole burst in all cases.}
    \label{Figx:fa time evolution}
\end{figure}

Finally, we investigated the $f_{a}$ time evolution of these 16 multi-peaked bursts in Figure \ref{Figx:fa time evolution} using the same order as in Figure \ref{Figurex:the light curve}.
The horizontal black dashed lines correspond to $f_{a}$ = 1.
In all multi-peaked bursts, the $f_{a}$ values vary in the range of $\sim1-7$ during the bursting period, which is similar to the range observed by \cite{Worpel2013ApJ} in other bursts ($f_{a}\sim2-10$).
In general, high values of $f_{a}$ coincide with large flux values during the burst.
We study the correlation between $f_{a}$ and bolometric flux (e.g. blackbody temperature and radius) for the 16 multi-peaked bursts using the method of cross-correlation lags \citep{Peterson1998PASP, Sun2018ascl.soft05032S}. 
We do not find any obvious correlation between $f_{a}$ and burst spectral parameters, and also we do not find any evidence of a relation between the centroid of the cross-correlation function (time delay between $f_{a}$ and bolometric flux) and the double-peaked structure (e.g. the peak-flux ratio).

 .

\subsection{The quadruple-peaked X-ray burst}
\label{The quadruple-peaked episode}

\begin{figure}
    \centering
        \includegraphics[height=3.5in,width=3.5in,angle=0]{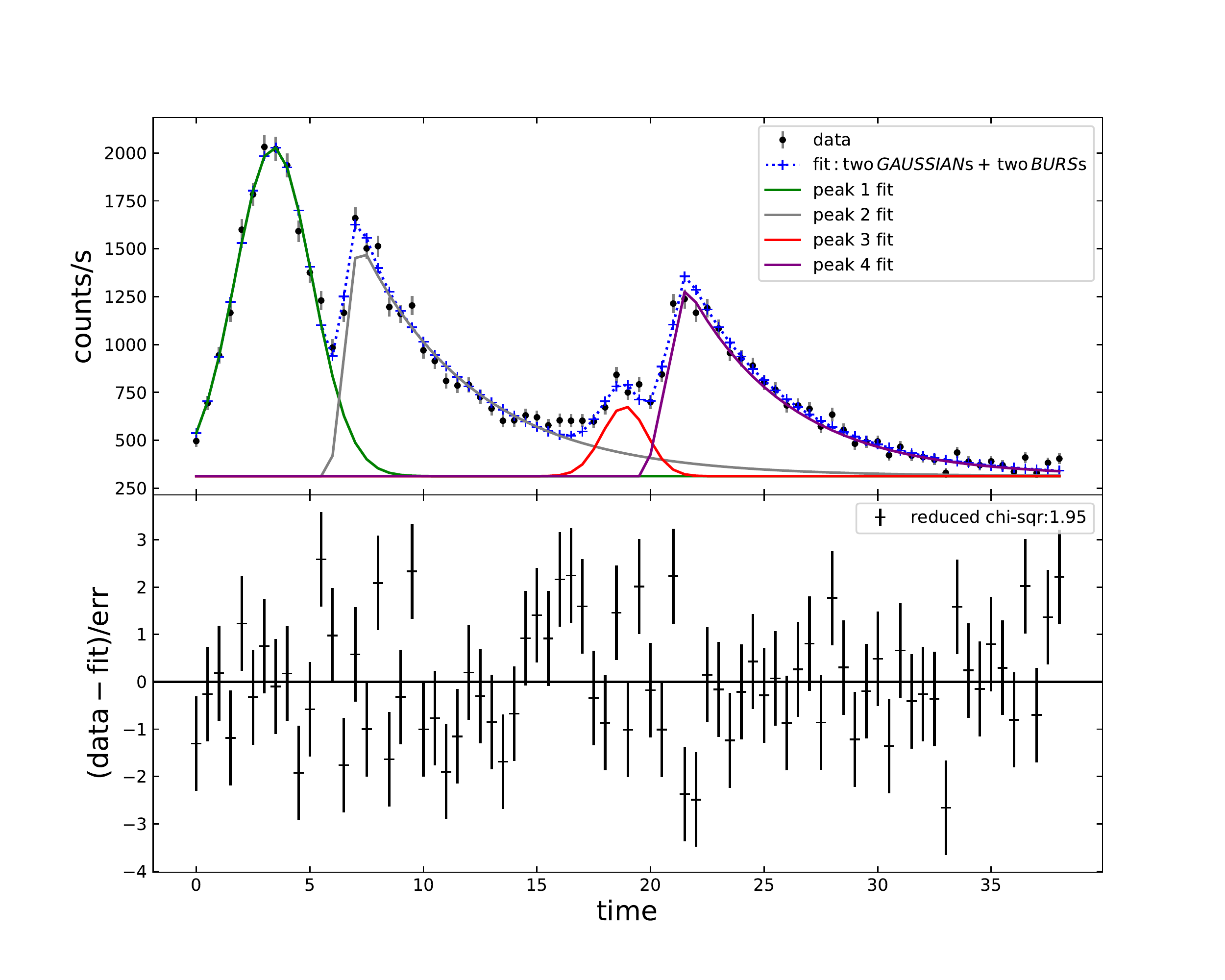}
    \caption{The light curve of the quadruple-peaked burst (burst $\#16$) from Obsid 95087-01-82-10 of 4U 1636$-$53. 
    The four peaks can be well described by two \textit{GAUSSIAN} functions adding two \textit{BURS} models in light curve.
   }
    \label{Fig: Obsid95087-01-82-10_light curve}
\end{figure}

In Figure \ref{Fig: Obsid95087-01-82-10_light curve}, we show the light curve of the quadruple-peaked X-ray burst. 
We use two \textit{GAUSSIAN} functions adding two \textit{BURS} models to fit the quadruple-peaked burst light curve data, where \textit{BURS} represents a model to describe the burst component in QDP \footnote[1]{https://heasarc.gsfc.nasa.gov/ftools/others/qdp/qdp.html}.


\begin{equation}
FNY=\left\{
\begin{array}{rcl}
0 && {t<ST} \\
BN*(t-ST)/(PT-ST) &&  {ST<t<PT} \\
BN*EXP(-(t-PT)/DT) && {PT<t} \\
\end{array}\right.,
\end{equation}
where $t$ is time, $FNY$ is the photons per second, $ST$ and $PT$ represent the start and peak time of each peak, respectively, and $BN$ and $DT$ stand for the normalisation and decay factor, respectively.

In the top panel of Figure \ref{Fig: Obsid95087-01-82-10_light curve}, we show the fitting results, getting a $\chi^2$ = 120.64 for 62 d.o.f. 
In the bottom, we show the residual of the fit in units of the error, and the black horizontal line represents the (data-fit)/err = 0.
To verify the existence of the third peak, we also use three components (one \textit{GAUSSIAN} function adding two \textit{BURS} models) to re-fit the light curve, getting a $\chi^2$ = 386.79 for 65 d.o.f. 
The F-test probability for these two fits is $1.1 \times 10^{-15}$, which indicates that the third peak is significant.
This is the first time that a quadruple-peaked burst is reported in 4U 1636$-$53.

There are three relatively strong peaks ($\sim 3 \rm s$, $\sim 7 \rm s$, $\sim 22 \rm s$ for the first, second, fourth peak) in the light curve and one weak peak ($\sim 19 \rm s$ for the third peak).
There is a very long time separation ($\sim12$s) between the second and the third peak.
The separation time between the first and last peak in this quadruple-peaked burst is $\sim 18$s, which is similar to the separation ($\sim17$s) in the triple-peaked burst reported by \cite{vanParadijs1986MNRAS}, but about two times longer than that in the triple-peaked burst ($\sim8$s) in \cite{Zhanggb2009MNRAS}.

\begin{figure}
    \centering
        \includegraphics[height=5.0in,width=3.5in,angle=0]{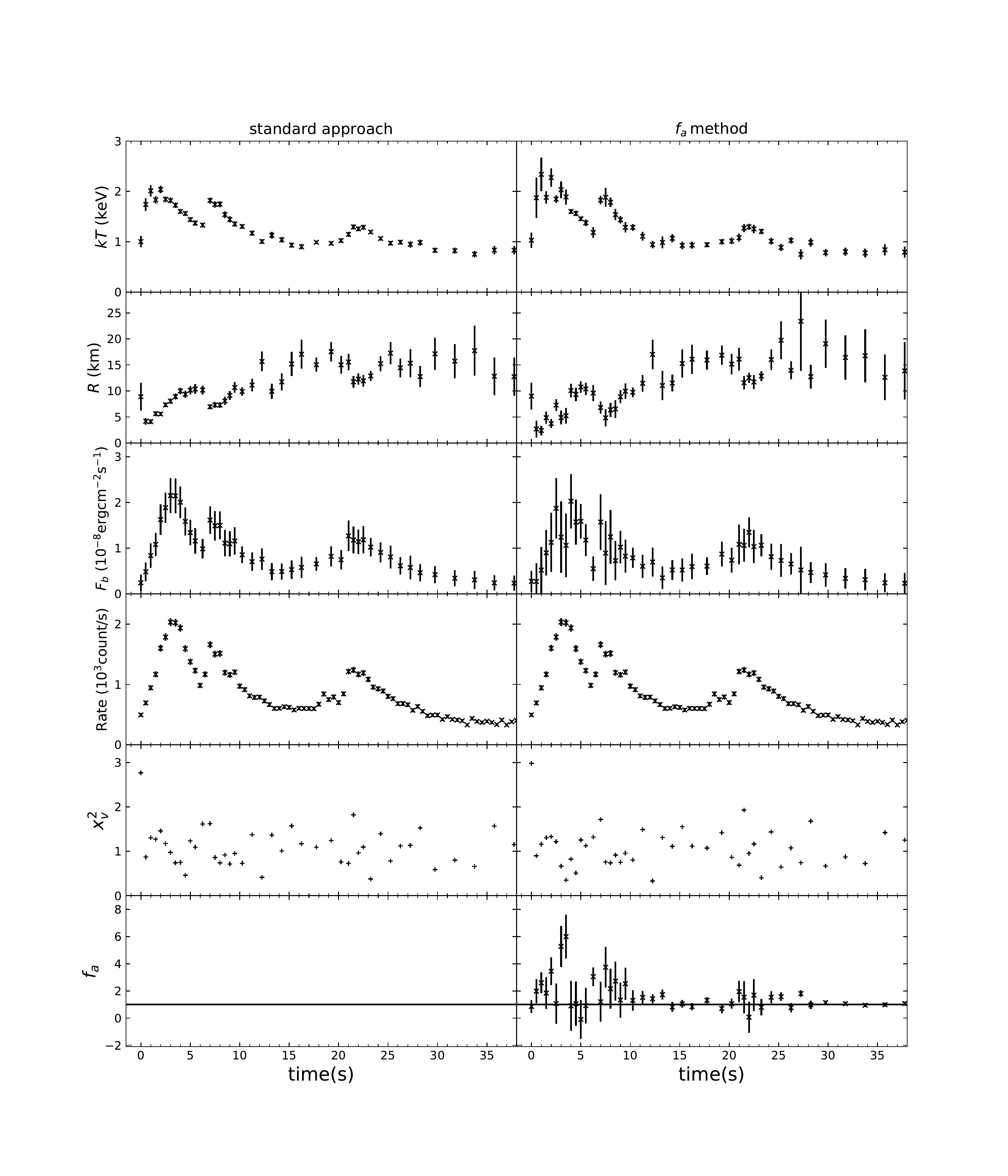}
    \caption{The spectral evolution of the quadruple-peaked burst (burst $\#16$) from Obsid 95087-01-82-10 of 4U 1636$-$53 using both the standard approach and the $f_{a}$ method. 
    The explanations of these panels are similar to footnote in the Figure \ref{Fig: Obsid60032-01-13-01}.
    The horizontal solid line in the bottom panel stands for $f_{a}$=1.
   }
    \label{Fig: Obsid95087-01-82-10}
\end{figure}

We investigated the time-resolved spectra of the quadruple-peaked burst with two models.
In the standard approach (the left-hand panel of Figure \ref{Fig: Obsid95087-01-82-10}), both the bolometric light curve and X-ray light curve show four peaks, three pronounced and small one.
The total burst fluence is $(26.9 \pm 3.5) \times 10^{-8} \; \rm ergs \; cm^{-2}$ and the bolometric first peak flux is $(2.0 \pm 0.6) \times 10^{-8} \; \rm ergs \; cm^{-2} \; s^{-1}$. 
There is a $\sim5$s long waiting period from 13s to 18s in the light curve,
and the bolometric flux is always higher than the persistent emission.
The effective temperature profile, after an initially rising, 
decreases with time and shows three noticeable local maxima. 
The blackbody radius increases steadily during the whole burst, which is similar to what was observed in the triple-peaked burst in \cite{Zhanggb2009MNRAS}.

In the $f_{a}$ method (the right-hand panel of Figure \ref{Fig: Obsid95087-01-82-10}), both the bolometric flux and X-ray light curve display four peaks as in the standard approach.
Similar to the double-peaked bursts, the bolometric flux in the $f_{a}$ method is lower than that of the standard approach due to the variation $f_{a}$ being larger than one during the first two peaks.
The evolution of the blackbody radius and the blackbody temperature in the $f_{a}$ method is similar to that in the standard approach.

\section{DISCUSSION}
\label{DISCUSSION}

We analyzed all available RXTE data of the LMXB 4U 1636$-$53, and we found 14 double-peaked bursts with complex profiles, one triple-peaked burst, and one quadruple-peaked burst that had not been reported before. 
The bolometric flux light curve of the multiple-peaked bursts exhibits a similar profile as the X-ray light curve, which indicates that the multi-peaked structures in 4U 1636$-$53 are not due to passband instrumental effect \citep{Jaisawal2019ApJ}. 
Between the peaks, the flux of these multi-peaked bursts never drops near or below the pre-burst level, which is different from the triple bursts in \cite{Boirin07} and \cite{Beri2019MNRAS}.

We find a marginal positive correlation between the peak-separation time and the peak-flux ratio 
in Class 1 bursts (bottom panel of Figure \ref{Figx:PD-PR}).
This suggests that double-peaked bursts with high peak-flux ratio value have a longer peak separation time.
This phenomenon may be explained if the burst consumes less fuel 
during the first peak, such that it is easier to release subsequently the fuel that is left  to trigger the second peak.
Therefore, the bursts with higher peak flux ratio need more time to accumulate fuel to release the second peak energy. 
As we know, in the bottom panel of Figure \ref{Fig:Flux1-Flux2_duration}, we also find an anti-correlation between the second peak flux and the separation time between two consecutive peaks. 
This indicates that the double-peaked bursts with longer peak separation time have a weaker second peak.  
If we assume that the mass accretion is negligible during the burst, this anti-correlation indicates that when the separation time is long, there is less fuel 
for the second peak.  
These results indicate that a single peak in these double-peaked bursts is not isolated, and the double-peaked profile is affected by the strength and separation time of the two single peaks.

Most of the multi-peaked bursts in our sample appear during the transition from the hard to the soft state in the CCD  where PRE bursts are also present \citep{Watts2007AA, Zhanggb2009MNRAS}. 
That the PRE and multi-peaked bursts appear in the same state of the source may provide an important clue to understanding the origin of the multi-peaked bursts, indicating that the appearance of the multi-peaked bursts could be affected by the transition of the source from the hard to the soft state.
If mass accretion rate increases from the upper right to the lower right in the CCD, there is no apparent relation between mass accretion rate and the parameter $r_{1,2}$.
We have checked the correlation between the persistent flux and the double-peaked burst parameters (peak ratio $r_{1,2}$; peak separation $\delta$), but do not find a clear trend between them either.

To further investigate if mass accretion affects the burst light curve structure, we studied the pre-burst spectra.  
We find that there is a positive correlation between the peak-flux ratio and the temperature of thermal component in the pre-burst spectra for the double-peaked bursts, where we use a blackbody model to fit the burst time-resolved spectra. 
The soft thermal component of X-ray spectra in accreting NS-LMXBs is generally explained by the emission from the NS surface and the accretion disc.
\cite{Sanna2013MNRAS} analyzed the X-ray spectra of six observations of 4U 1636$-$53 in different spectral states taken with XMM-Newton and RXTE simultaneously. 
They find that the temperature at the inner disc radius is $\sim 0.1-0.8$ keV, and the temperature at the NS surface is $\sim 1.4-2.0$ keV. 
We used a black-body or a disk black-body to depict the enhancement of the thermal component. The emission from the NS surface and inner region of the accretion disc are degenerate in our persistent spectrum. 
The high blackbody (or disc blackbody) temperature could be due to an enhanced accretion rate in the disc or an increased temperature of the NS surface. Since the above discussion indicates that the double-peaked properties are not affected by accretion rate, we suggest that the double-
peaked bursts with high peak flux ratio might appear when the NS surface temperature is high.


To date, there are several models to explain the unusual double-peaked structure in X-ray bursts.
One of them is the thermonuclear flame spreading model \citep{Bhattacharyya200601, Bhattacharyya200604}, 
which succeeded in explaining the large dip in the X-ray light curve and reproducing the spectral evolution of double-peaked bursts. 
\cite{Cooper07} suggested that ignition latitude has a positive correlation with the accretion rate. 
Combing these ideas with the results in Figure \ref{Figx:pre-par_PR} and assuming that the temperature of the NS is correlated to the accretion rate, the high peak flux ratio of double-peaked bursts would correspond to high latitude ignition \citep{Bhattacharyya200601,Bhattacharyya200604}.

Alternatively, \cite{Lampe2016} investigated the possibility of the nuclear origin of the double-peaked bursts based on the model of nuclear waiting points in the rp-process explaining the double-peaked profile \citep{Fisker04}. 
They find that for certain metallicities and low accretion rate, there is an anti-correlation between the peak-flux ratio ($r_{1,2}$) and accretion rate.
We do not find that the peak-flux ratio of double-peaked burst decreases when the source moves in the CCD (see Figure \ref{Figx:CCD}) from the top right to the bottom right, as the accretion rate gradually increases.
If the temperature before the burst is propotional to mass accretion rate, the correlation between the peak flux and the accretion rate does not agree (see Figure \ref{Figx:pre-par_PR}) with the simulation of \cite{Lampe2016} .
In their simulation, as accretion rate increases, the double-peaked structure shows a distinct two stage, different from the large dip in the bolometric flux profile of our sample (See Figure \ref{Figurex:the light curve}).

The newly discovered quadruple-peaked burst in 4U 1636$-$53 poses a problem to the thermonuclear spreading model of \cite{Bhattacharyya200601,Bhattacharyya200604}. 
If this model is applied to explain the four peaks, the burning front not only needs to stall three time, but it also needs a longer "waiting" period between the second and third peak than the separation time in the double-peaked burst.

Similar to normal single-peaked bursts, we find that the $f_{a}$ value is large when the flux is high during the multi-peaked bursts.
We investigated the relation between $f_{a}$ and the burst spectral parameters, but do not find any obvious correlation between them. 
The enhanced $f_{a}$ can be interpreted as changes in the accretion flow rate by Poynting-Robertson drag \citep{Walker1992ApJ}. 
However, it is difficult to explain the dip between peaks by reducing the accretion.

The double-peaked structure could be due to the influence of the variation of the accretion geometry, but we should observe more double-peaked bursts to test this hypothesis. We note also that it is hard to explain the existence of triple or quadruple-peaked profiles  with the variation of the geometry. 


\section{SUMMARY}
\label{SUMMARY}

We found 16 bursts with multi-peaked structure by investigating 336 type I X-ray bursts in LMXB 4U 1636$-$53 with RXTE. Our sample contains 14 double-peaked, one triple-peaked and one quadruple-peaked burst; the latter had never been reported before.

(i) Most of the multi-peaked bursts in our sample appear during the transition from the hard to the soft state in the CCD.

(ii) We find that double-peaked bursts with high peak-flux ratio appear when the pre-burst temperature is high; the high NS temperature may be due to enhanced accretion rate on to the NS surface.

(iii) We find an anti-correlation between the second peak flux and the peak-separation time for double-peaked bursts.

(iv) The quadruple-peaked burst shows a long separation time between the second and the third peak which is difficult to explain with current models. 

(v) We use the $f_{a}$ method to re-analyse these 16 bursts and we find no evidence that the multi-peaked structure is due to enhanced accretion during the bursts.

\section*{ACKNOWLEDGEMENTS}
\label{ACKNOWLEDGEMENTS}
This research has made use of data obtained from the High Energy Astrophysics Science Archive Research Center (HEASARC), provided by NASA’s Goddard Space Flight Center and NASA’s Astrophysics Data System Bibliographic Services.
We thank Liantao Cheng, Shasha Li for their helpful discussions.
G.B. acknowledges funding support from the National Natural Science Foundation of China (NSFC) under grant No. U1838116 and Y7CZ181002.
M.L. is supported by National Natural Science Foundation of China (grant No.11803025) and the Hunan Provincial Natural Science Foundation (grant No. 2018JJ3483) .




\bibliographystyle{mnras}
\bibliography{wenxian} 



\appendix



\section{Double-peaked events in our sample}
\label{Double-peaked events in our sample}

Some bursts appear to be flat after the onset followed by a clear peak, such as burst $\#$6 and $\#$8. In order to verify the significance of the double-peaked structure of these bursts, we use either one or two Gaussian functions to fit the bursts around the peaking time data.

\begin{figure}
\centering
\includegraphics[height=1.6in,width=1.6in,angle=0]{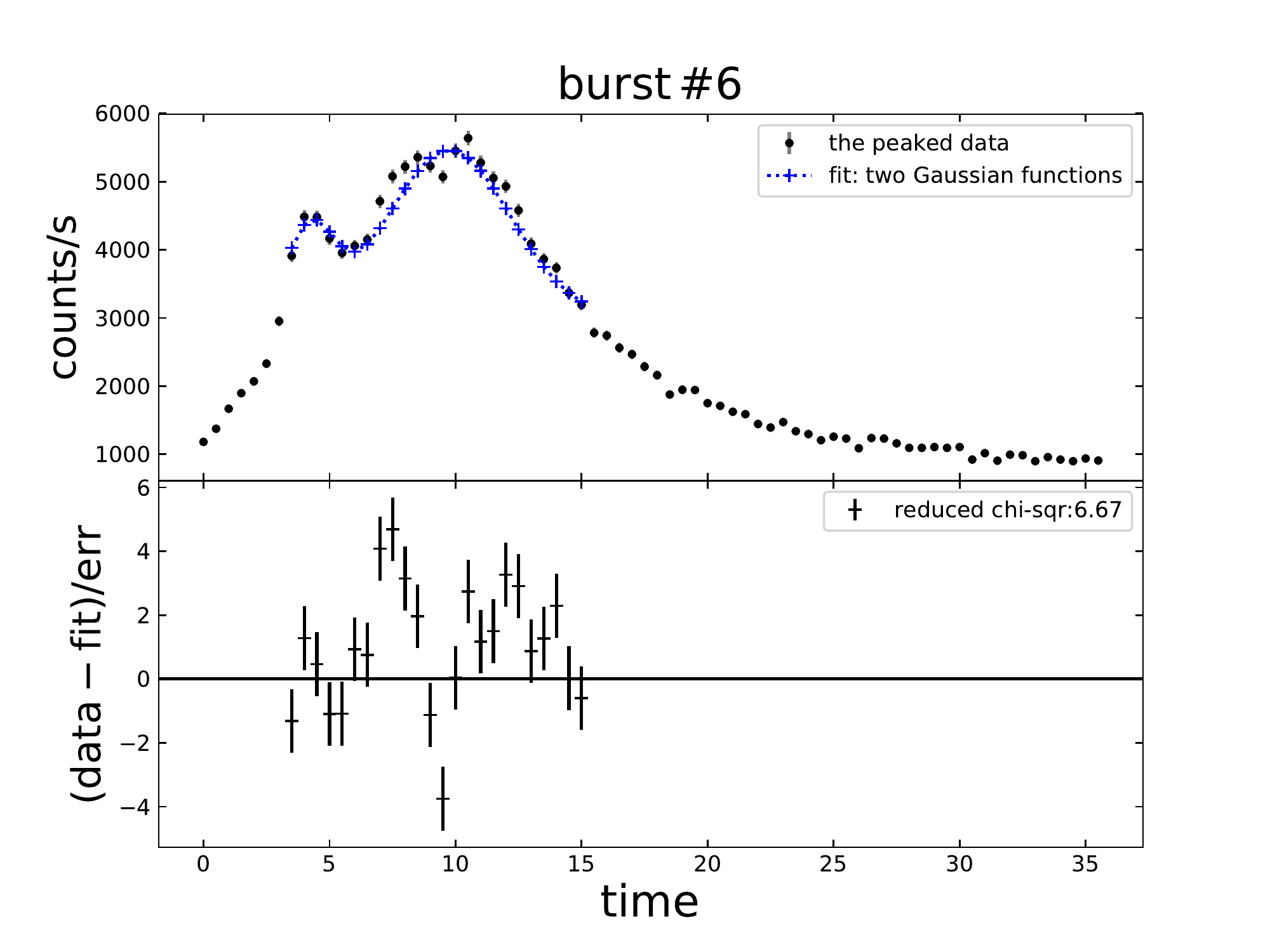}
\includegraphics[height=1.6in,width=1.6in,angle=0]{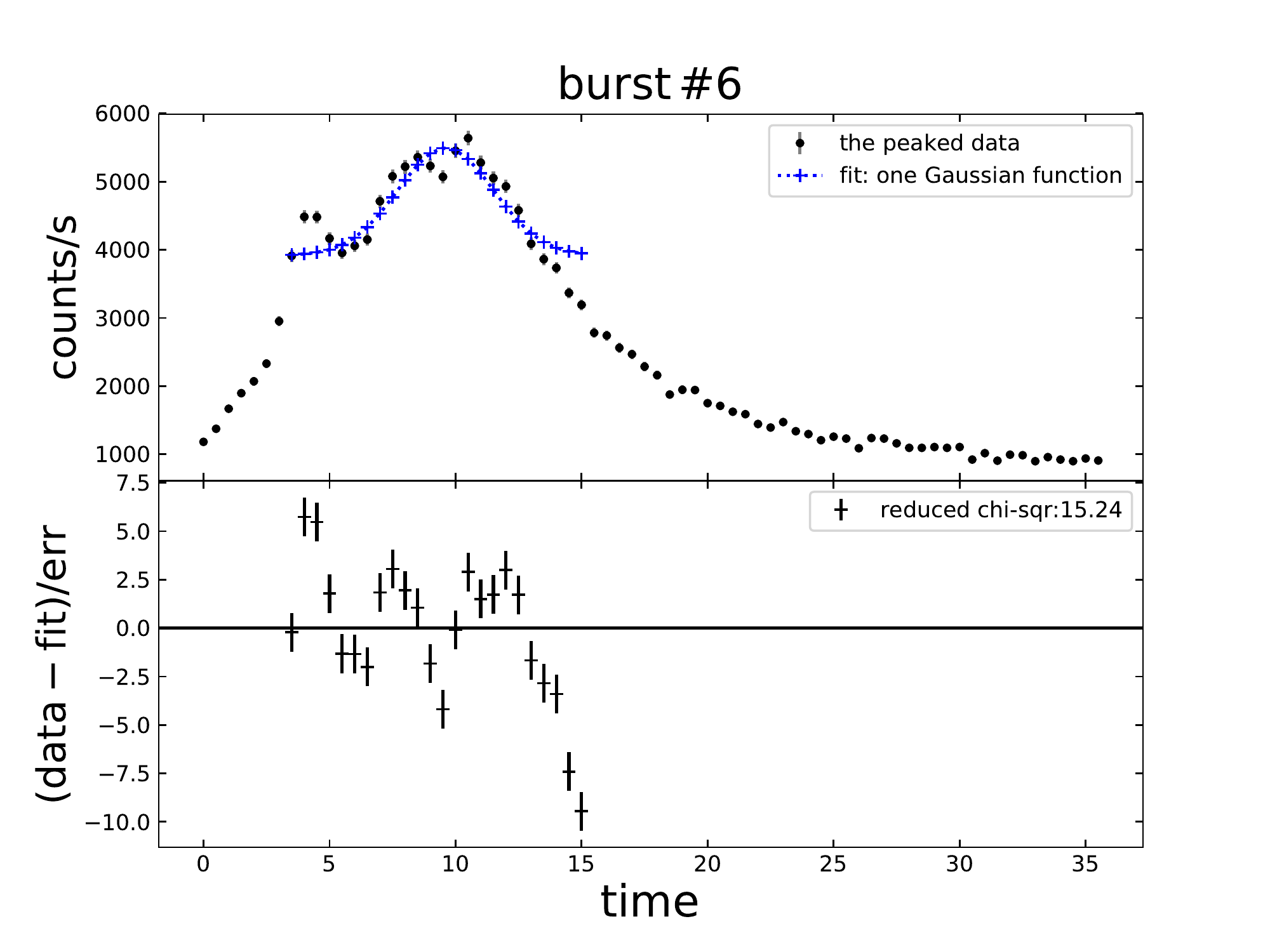}
\includegraphics[height=1.5in,width=1.6in,angle=0]{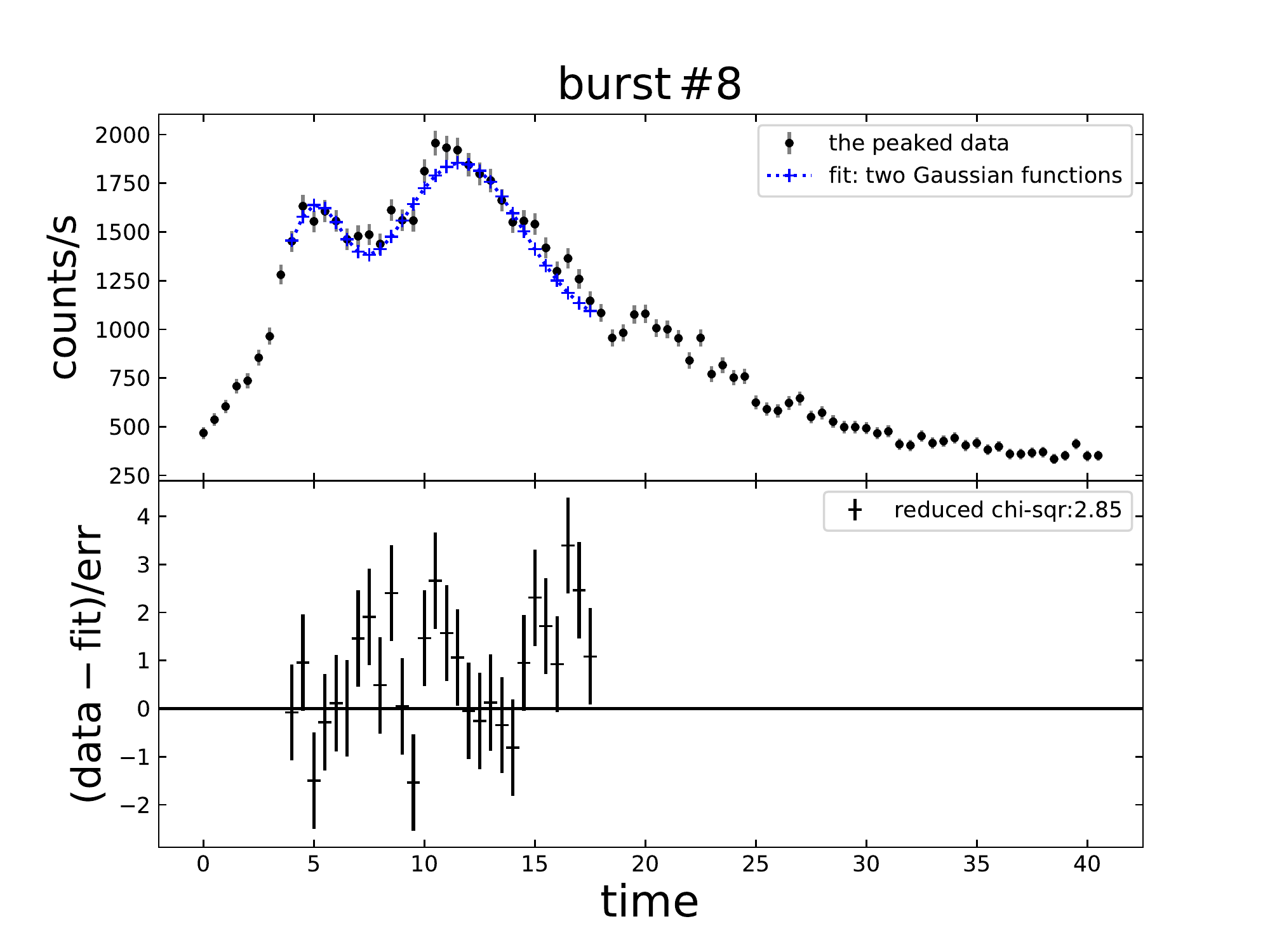}
\includegraphics[height=1.5in,width=1.6in,angle=0]{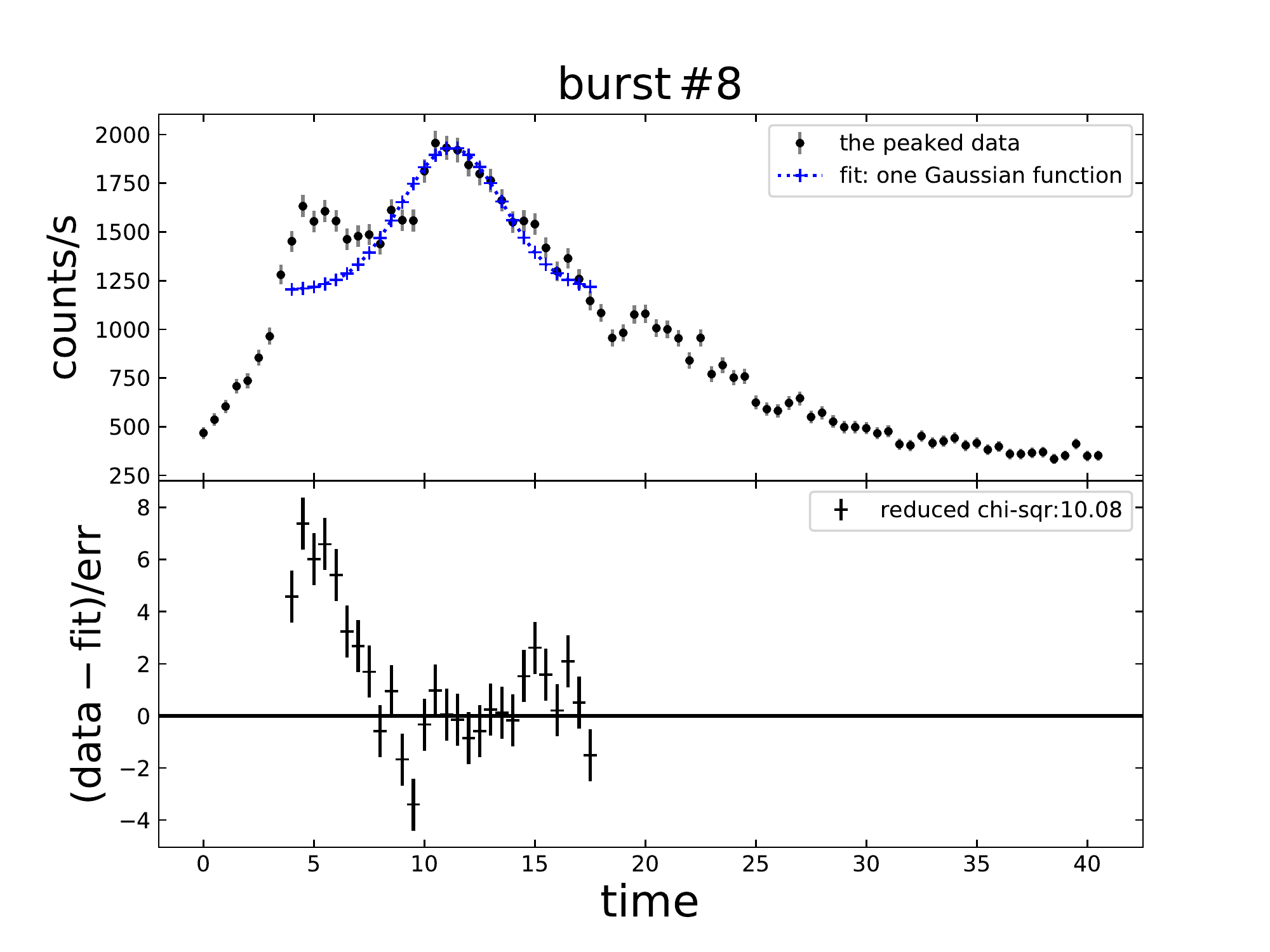}
\caption{
The light curves of burst $\#6$ and $\#8$ of 4U 1636$-$53. The peaking time data of these two bursts can be well described by two Gaussian function.
}
\label{Figx:burst6_8}
\end{figure}

For burst $\#$6, at the top left panel of \ref{Figx:burst6_8}, we show the fitting results of the two Gaussian functions, yielding $\chi^2 = 113.3$ for 17 d.o.f, and at the top right panel of \ref{Figx:burst6_8}, we show the fitting results with one Gaussian function, yielding $\chi^2 = 304.8$ for 20 d.o.f. The F-test probability for these two fits is $0.0006$, which indicates that the double-peaked structure is significant.
 
For burst $\#$8, at the bottom left panel of \ref{Figx:burst6_8}, we show the fitting results of the two Gaussian functions, we yielding $\chi^2 = 59.8$ for 21 d.o.f, and at the bottom right panel of \ref{Figx:burst6_8}, we show the fitting results with one Gaussian function, yielding $\chi^2 = 241.9$ for 24 d.o.f. The F-test probability for these two fits is $1.4 \times 10^{-6}$, which also indicates that the double-peaked structure is significant.

\bsp	
\label{lastpage}
\end{document}